\documentclass[12pt]{article}
\usepackage{amsmath,amsfonts}
\advance \textheight by \topskip
\makeatletter

\def\cc#1{\kern .7em\hfill #1 \hfill\kern .7em}

\newcommand{\beq}{\begin{equation}}
\newcommand{\bee}{\end{equation}}
\newcommand{\beqa}{\begin{eqnarray}}
\newcommand{\eeqa}{\end{eqnarray}}
\newcommand{\noi}{\noindent}
\newcommand{\e}{\varepsilon}
\newcommand{\ttA}{{\tilde {\tilde A}}}
\newcommand{\tA}{\tilde  A}
\newcommand{\ttB}{{\tilde {\tilde B}}}
\newcommand{\tB}{\tilde  B}
\newcommand{\ttP}{{\tilde {\tilde \varphi}}}
 \newcommand{\tP}{\tilde  \varphi}
\newcommand{\hhA}{{\hat {\hat A}}}
\newcommand{\hA}{\hat  A}
\newcommand{\hhB}{{\hat {\hat B}}}
\newcommand{\hB}{\hat  B}
\newcommand{\hhP}{{\hat {\hat \varphi}}}
 \newcommand{\hP}{\hat  \varphi}
\newcommand{\tpsi}{\tilde \psi}
\newcommand{\ttpsi}{{\tilde{ \tilde \psi}}}
\newcommand{\ttl}{{\tilde{\tilde  \lambda}}}
\newcommand{\tl}{\tilde \lambda}
\newcommand{\tr}{\tilde \rho}
\newcommand{\ttr}{{\tilde{\tilde  \rho}}}
\newcommand{\ttf}{{\tilde{\tilde  f}}}
\newcommand{\tf}{\tilde f}

\def\>{\rangle}
\def\<{\langle}

\begin{document}

\title{
\begin{flushright}
{\small PM/04-19}\\[.5cm]
\end{flushright}
{\bf Cubic supersymmetry and abelian gauge invariance} }
 

\author{    
{\sf   G. Moultaka} \thanks{e-mail:
moultaka@lpm.univ-montp2.fr}$\,\,$${}^{a},$
{\sf  M. Rausch de Traubenberg }\thanks{e-mail:
rausch@lpt1.u-strasbg.fr}$\,\,$${}^{b}$
 and
{\sf A. Tanas\u a}\thanks{e-mail:
atanasa@lpt1.u-strasbg.fr}$\,\,$$^{c,b}$\\
\\
{\small ${}^{a}${\it Laboratoire de Physique 
Math\'ematique et Th\'eorique, CNRS UMR 5825, 
Universit\'e Montpellier II,}}\\
{\small {\it Place E. Bataillon, 34095 Montpellier,
France}}\\
{\small ${}^{b}${\it
Laboratoire de Physique Th\'eorique, CNRS UMR  7085,
Universit\'e Louis Pasteur}}\\
{\small {\it  3 rue de 
l'Universit\'e, 67084 Strasbourg, France}}\\ 
{\small ${}^{c}${\it
Laboratoire de Math\'ematique et Applications,
Universit\'e de Haute Alsace }}\\
{\small {\it  Facult\'e des Sciences et Techniques, 4 rue des 
fr\`eres Lumi\`eres 68093  Mulhouse, France}} 
}
\date{}
\maketitle
\vskip-1.5cm

\vspace{2truecm}

\begin{abstract}
\noindent
On the basis of recent results extending  non-trivially the Poincar\'e
symmetry, we investigate  the properties of bosonic 
multiplets including $2-$form gauge fields. Invariant free 
Lagrangians are explicitly built which involve possibly   $3-$ and 
$4-$form fields. We also study in detail the interplay between this
symmetry and a $U(1)$ gauge symmetry, and in particular the implications
of the automatic gauge-fixing of the latter associated to a residual gauge 
invariance, as well as the absence of self-interaction terms.  
\end{abstract}

\vspace{2cm}

keywords: extensions of the Poincar\'e algebra, Field Theory,
algebraic methods, Lie (super)algebras, gauge symmetry.

\newpage

\section{Introduction}
\renewcommand{\theequation}{1.\arabic{equation}}   
\setcounter{equation}{0}
A non-trivial extension of the Poincar\'e algebra, different from the 
supersymmetric one, was introduced in  \cite{flie}. The main idea
is to consider an adapted algebraic structure, named $F-$Lie
algebra, which is a generalisation of Lie superalgebras.
Consequently, from the very beginning  
this construction evades the no-go theorem of 
Haag-Lopuszanski-Sohnius \cite{hls}.
Similarly to Lie superalgebras which underly the structure of supersymmetry,
$F-$Lie algebras underly that of fractional
supersymmetry  \cite{fsusy}.

A specific $F-$Lie algebra (for $F=3$) has been studied
and leads to 
a quantum field  theoretical  realisation  of a non-interacting 
theory, named  {\it cubic supersymmetry} or 3SUSY \cite{cubic}.
In this new algebraic frame,  one  does not consider
square roots of translations ($QQ \sim P$), as it is the case 
for supersymmetry, but rather cubic roots ($QQQ \sim P$).
The representation theory of 3SUSY has been investigated and leads
either to pure bosonic or to pure fermionic multiplets. The
situation is drastically different from supersymmetry, since
the multiplets contain only states with the same  statistics. This is
due to the fact that in our algebra, the additional generators
$Q$  belong to   the vector representation of the Poincar\'e algebra,
while in the SUSY case the additional generators  
belong  to  the spinorial 
representation of the Poincar\'e algebra.

In this paper we investigate the properties of the bosonic multiplets
which involve scalar, vectors and $2-$forms.
In the section $2$ we firstly recall some basic results already obtained
in \cite{cubic}.  Then, we explicitly diagonalise the 
Lagrangian obtained in \cite{cubic}. 
We observe that 3SUSY invariance requires
gauge fixing terms {\it \`a la } Feynman,
 for the vectors and the $2-$forms.
This Lagrangian has wrong signs in  the kinetic term 
of some of the fields, thus  leading
to unboundedness from below for these energy densities. We propose
here a possible solution to this problem. Indeed, using the Hodge duality
for the $p-$forms and the specific form of the Lagrangian (kinetic term +
gauge fixing term), the physical field is interpreted as ${}^\star A$, the
Hodge dual of $A$, instead of $A$. This mechanism leads to
$3-$ and $4-$forms. Then, quadratic couplings between different 
types of bosonic  mutiplets are taken into account.  The total
free Lagrangian is then diagonalised, leading to (i)  constraints on
the coupling  parameters in order to have positive square mass,  and  
(ii) non-conventional kinetic terms for the $2-$forms. 
Section $3$ is devoted to the proof that no 3SUSY invariant interacting
terms are possible within these bosonic multiplets. In this section we 
also recall some relations satisfied by the (anti)-self-dual 2-froms,  
and establish a useful property for the derivatives of the various
multiplets. In section $4$, we study the compatibility between the 3SUSY
and $U(1)$ (gauge) symmetries, point out the existence of an induced symmetry, 
and determine explicitly the functional subclass of the allowed gauge 
transformations. We also comment briefly on tentative superspace formulation.  
Section $5$ contains the conclusions and some perspectives as regards
the interaction possibilities.

\section{Free theory}
\renewcommand{\theequation}{2.\arabic{equation}}   
\setcounter{equation}{0}
\subsection{Algebra and self-coupling of multiplets}

The 3SUSY algebra is constructed from the Poincar\'e generators
$P_m, L_{mn}$ with additional generators $Q_m$ in the vector representation
of the Lorentz group \cite{cubic}

 \beqa
\label{algebra}
&&\left[L_{mn}, L_{pq}\right]=
\eta_{nq} L_{pm}-\eta_{mq} L_{pn} + \eta_{np}L_{mq}-\eta_{mp} L_{nq},
\ \left[L_{mn}, P_p \right]= \eta_{np} P_m -\eta_{mp} P_n, \nonumber \\
&&\left[L_{mn}, Q_p \right]= \eta_{np} Q_m -\eta_{mp} Q_n, \ \
\left[P_{m}, Q_n \right]= 0, \\
&&\left\{Q_m, Q_n, Q_r \right \}=
\eta_{m n} P_r +  \eta_{m r} P_n + \eta_{r n} P_m, \nonumber
\eeqa

 \noindent
where $\{Q_m,Q_n,Q_p \}=
Q_m Q_n Q_r + Q_m Q_r Q_n + Q_n Q_m Q_r + Q_n Q_r Q_m + Q_r Q_m Q_n +
Q_r Q_n Q_m $  stands for the symmetric product of order $3$ and
$\eta_{mn} = \mathrm{diag}\left(1,-1,-1,-1\right)$  is the Minkowski
metric.  Two irreducible matrix representations  have been found

\begin{eqnarray}
\label{matirred}
Q_{+}{}_m=\begin{pmatrix} 0&\Lambda^{1/3} \sigma_m& 0 \cr
                           0&0&\Lambda^{1/3}\bar \sigma_m \cr
                           \Lambda^{-2/3}P_m&0&0
\end{pmatrix},
Q_{-}{}_m=\begin{pmatrix} 0&\Lambda^{1/3} \bar \sigma_m& 0 \cr
                           0&0&\Lambda^{1/3}\sigma_m \cr
                           \Lambda^{-2/3}P_m&0&0
\end{pmatrix}
\end{eqnarray}

\noi
with   $\sigma^m=(\sigma^0=1,\sigma^i)$, and 
$\bar \sigma^m =(\bar \sigma^0=1,-\sigma^i)$,  $\sigma^i$ 
the Pauli matrices
and $\Lambda$ a 
parameter with mass dimension   that we take  equal to $1$ 
(in appropriate units).  These matrix representations give rise 
to fermionic and bosonic multiplets \cite{cubic}. Here,
we just consider the following bosonic multiplets \cite{cubic}:

\beqa
\label{4-decomposition}
\Xi_{++}=
 \begin{pmatrix} 
\varphi, B_{mn} \cr \tA_m \cr \ttP, \ttB_{mn}
\end{pmatrix}&& \ \ \ 
\Xi_{+-}=
\begin{pmatrix}    A'_m \cr \tP', \tB'_{mn} \cr  \ttA'_m
\end{pmatrix}
\nonumber 
\\
\Xi_{--}=
\begin{pmatrix}
\varphi', B^\prime_{mn} \cr \tA'_m \cr \ttP', \ttB'_{mn}\end{pmatrix}
&& \ \ \ 
\Xi_{-+}=
\begin{pmatrix} A_m \cr \tP, \tB_{mn} \cr  \ttA_m \end{pmatrix}
\eeqa

\noi
where  $\varphi, \ttP, \varphi', \ttP',
\tP, \tP'$ are scalars fields, $\tA, \tA', A, \ttA, A',\ttA'$ are vector 
fields, $B, \tB, \ttB$ are self-dual $2-$forms
and $B',\tB', \ttB'$ are anti-self-dual $2-$forms. The 3SUSY 
algebra  (\ref{algebra})  
and its representations  (\ref{matirred}) are $\mathbb Z_3-$graded.
Therefore, one can assume that, for example, 
for the multiplet $\Xi_{++}$, the fields $\varphi, B$
are in the $(-1)-$graded sector, $\tA$ in the $0-$graded sector
and $\ttP, \ttB$ in the $1-$graded sector. The same classification
also holds for the other multiplets. Furthermore, 
due  to the property of (anti-)self-duality  of  $2-$forms in
 $4D$, ${}^\star B= i B,  {}^\star B'= -i B'$, {\it etc} 
(with $ {}^\star B$ the
Hodge dual of $B$), the $2-$forms are 
complex representations of
$\mathfrak{so}(1,3)$ and consequently also the scalars and vector fields
(see Eq.[\ref{transfo2}] below). 
These multiplets have been obtained from the matrices 
$Q_\pm$, with the vacua in the spinor representations of the
Lorentz algebra \cite{cubic}. For instance,  we have
$\Xi_{++}= \begin{pmatrix} \Psi_1{}_{ +} \cr
\Bar \Psi_2{}_{-} \cr \Psi_3{}_{+} \end{pmatrix} \otimes \Omega_+$
with $\Psi_1{}_{+}, \Psi_3{}_{+}$ two left-handed spinors,
$\Bar \Psi_2{}_{-}$ a right handed spinor and $\Omega_+$, the
vacuum,  a left-handed spinor. The transformation law
for $\Xi_{++}$ is then obtained from

\beqa
\label{transfo}
 \delta_\e \Xi_{++} =
\left(\e^m Q_{+m} \begin{pmatrix} \Psi_1{}_{+} \cr
\Bar \Psi_2{}_{-} \cr \Psi_3{}_{+} \end{pmatrix} \right)
\otimes \Omega_+
\eeqa

\noi
Similar definitions hold for the three other multiplets. We recall here
the corresponding transformation laws obtained after some algebraic 
manipulation \cite{cubic}

\beqa
\label{transfo2}
 \begin{array}{ll}
~~~~~~~~~~~~~~~~~~~~~~~~~{}^{(+,+)} & ~~~~~~~~~~~~~~~~~~~~~~~~~{}^{(+,-)} \cr
 \left\{\begin{array}{l}
\delta_\varepsilon \varphi =  
\varepsilon^m \tA_m   \cr
\delta_\varepsilon B_{mn} = - (
\varepsilon_m \tA_n - \varepsilon_n \tA_m )
+ i \varepsilon_{mnpq} \varepsilon^p \tA^q{} \cr
\delta_\varepsilon \tA_m =  (
\varepsilon_m \ttP +
\varepsilon^n  \ttB_{mn} ) \cr
\delta_\varepsilon \ttP = \varepsilon^m
\partial_m \varphi \ \ \  
\delta_\varepsilon  \ttB_{mn} = \varepsilon^p \partial_p
B_{mn} \end{array} \right. &
\left\{ \begin{array}{l}
\delta_\varepsilon A_m^\prime = (\varepsilon^n \tB^\prime_{mn}+
\varepsilon_m \tP^\prime) \cr
\delta_\varepsilon \tP^\prime =  \varepsilon^m 
\ttA_m^\prime \cr
\delta_\varepsilon \tB^\prime _{mn} = - (
\varepsilon_m \ttA^\prime_n - \varepsilon_n \ttA^\prime_m ) 
- i \varepsilon_{mnpq} \varepsilon^p \ttA^\prime{}^q{}
\cr
\delta \ttA^\prime_m =  \varepsilon^n \partial_n A^\prime_m
\end{array}\right. \end{array} 
\nonumber \\
\\
\begin{array}{ll}
~~~~~~~~~~~~~~~~~~~~~~~~~{}^{(-,-)} & ~~~~~~~~~~~~~~~~~~~~~~~~~{}^{(-,+)} \cr
\left\{\begin{array}{l}
\delta_\varepsilon \varphi' =  
\varepsilon^m \tA'_m   \cr
\delta_\varepsilon B'_{mn} = - (
\varepsilon_m \tA'_n - \varepsilon_n \tA'_m )
- i \varepsilon_{mnpq} \varepsilon^p \tA'^q{} \cr
\delta_\varepsilon \tA'_m =  (
\varepsilon_m \ttP' +
\varepsilon^n  \ttB'_{mn} ) \cr
\delta_\varepsilon \ttP' = \varepsilon^m
\partial_m \varphi' \ ,  \ \ \  
\delta_\varepsilon  \ttB'_{mn} = \varepsilon^p \partial_p
B'_{mn} \end{array} \right. &
\left\{ \begin{array}{l}
\delta_\varepsilon A_m = (\varepsilon^n \tB_{mn}+
\varepsilon_m \tP) \cr
\delta_\varepsilon \tP =  \varepsilon^m 
\ttA_m \cr
\delta_\varepsilon \tB _{mn} = - (
\varepsilon_m \ttA_n - \varepsilon_n \ttA_m ) 
+ i \varepsilon_{mnpq} \varepsilon^p \ttA{}^q{}
\cr
\delta \ttA_m =  \varepsilon^n \partial_n A_m
\end{array}\right. \end{array} \nonumber 
\eeqa

\noi%
with $\e$ a real Lorentz vector and $P_m = \partial_m$
(this is a slight difference compared to \cite{cubic}, where
$\e$ was taken purely imaginary). As can be seen from the transformation
laws (\ref{transfo2}), the complex conjugate of $\Xi_{--}$ (resp.
$\Xi_{-+}$) transforms like $\Xi_{++}$ (resp. $\Xi_{+-}$).
In the following we will thus consider the minimal set of field content,
taking $\Xi_{++}^\star= \Xi_{--}, \Xi_{+-}^\star = \Xi_{-+}$ ({\sl i.e.}
$\varphi^\star = \varphi', \tA^\star = \tA'$ {\it etc.} ), so that the 
multiplet $\Xi_{ab}$ is the CPT conjugate of  $\Xi_{-a -b}$.\\

We introduce for each $1-$form potential $A_m$ the $2-$form  field strength 
$F_{mn}=\partial_m A_n -\partial_n A_m$, and for each $2-$form potential a
$3-$form field strength 
$H_{mnp}=\partial_m B_{np} + \partial_n B_{pm} + \partial_p B_{mn}$
together with its  dual $1-$form 
${}^\star H_m=\frac16 \varepsilon_{mnpq}H^{npq}$. 
We can construct  two zero-graded real 3SUSY invariant
Lagrangians associated respectively to
the multiplets $(\Xi_{++}, \Xi_{--})$ and $(\Xi_{+-}, \Xi_{-+})$
\cite{cubic}

\beqa
\label{free4}
{\cal L}_0&=&{\cal L}_0(\Xi_{++}) +{\cal L}_0(\Xi_{--} ) \nonumber \\
&=&\partial_m \varphi \partial^m \ttP + \frac{1}{12} H_{mnp}
{\tilde {\tilde H}}^{mnp} -\frac12 {}^\star H_m {}^\star {\tilde {\tilde
H}}^m -\frac14 \tilde F_{mn} \tilde F^{mn} -
\frac12 \left(\partial_m \tilde A^m\right)^2 \nonumber \\
&+& \partial_m \varphi' \partial^m \ttP' + \frac{1}{12} H'_{mnp}
{\tilde {\tilde H'}}^{mnp} -\frac12 {}^\star H'_m {}^\star {\tilde {\tilde
H'}}^m -\frac14 \tilde F'_{mn} \tilde F'^{mn} -
\frac12 \left(\partial_m \tilde A'^m\right)^2
\nonumber \\
\\
{\cal L'}_0&=&{\cal L}_0(\Xi_{-+}) +{\cal L}_0(\Xi_{+-} ) \nonumber \\
&=&\frac12 \partial_m \tP \partial^m \tP
+\frac{1}{24} \tilde H_{mnp} \tilde H^{mnp} -
\frac14  {}^\star \tilde H_m {}^\star \tilde H^m 
-\frac12 F_{mn} {\tilde {\tilde F}}^{mn} - 
(\partial_m A^m)(\partial_n \ttA^n)  \nonumber \\
&+&\frac12 \partial_m \tP' \partial^m \tP'
+\frac{1}{24} \tilde H'_{mnp} \tilde H'^{mnp} -
\frac14  {}^\star \tilde H'_m {}^\star \tilde H'^m 
-\frac12 F'_{mn} {\tilde {\tilde F'}}^{mn} - 
(\partial_m A'^m)(\partial_n \ttA'^n)  \nonumber
\eeqa

To identify the physical degrees of freedom, we proceed in several steps.
We concentrate just  on ${\cal L}_0$ {\it i.e} on the
multiplets $\Xi_{++}$ and $\Xi_{--}$. For ${\cal L'}_0$ the
results are analogous.

Firstly, we introduce the real fields

\beqa
\label{real}
\tA_1=\frac{\tA+\tA^\prime}{\sqrt{2}} &,&
\tA_2=  i\frac{\tA-\tA^\prime}{\sqrt{2}}, \nonumber \\
B_1= \frac{B+B^\prime}{\sqrt{2}} &,& 
B_2= i\frac{B-B^\prime}{\sqrt{2}}, \nonumber \\
\ttB_1=  \frac{\ttB+\ttB^\prime}{\sqrt{2}} &,& 
\ttB_2=  i\frac{\ttB- \ttB^\prime}{\sqrt{2}}, \\
\varphi_1= \frac{\varphi+\varphi^\prime}{\sqrt{2}} &,& 
\varphi_2= i\frac{\varphi-\varphi^\prime}{\sqrt{2}}, \nonumber \\
\ttP_1=  \frac{\ttP+\ttP^\prime}{\sqrt{2}} &,& 
\ttP_2=  i\frac{\ttP- \ttP^\prime}{\sqrt{2}}. \nonumber
\eeqa 

\noi
Then, using 

\beqa
\label{B1-B2}
{}^\star B_1 = B_2, {}^\star \ttB_1 = \ttB_2,
\eeqa

\noindent the Lagrangian ${\cal L}_0$ becomes

\beqa
\label{free4_2}
{\cal L}_0
&=&\partial_m \varphi_1 \partial^m \ttP_1 -
 \partial_m \varphi_2 \partial^m \ttP_2 
+ \frac{1}{6} H_1{}_{mnp} {\tilde {\tilde H}}_1^{mnp} 
+ \partial^n B_1{}_{nm} \partial_p \ttB_1{}^{pm} 
\nonumber \\
&-&\frac14 \tilde F_1{}_{mn} \tilde F_1{}^{mn} 
+\frac14 \tilde F_2{}_{mn} \tilde F_2{}^{mn} 
-\frac12 \left(\partial_m \tilde A_1{}^m\right)^2 
+\frac12 \left(\partial_m \tilde A_2{}^m\right)^2.
\eeqa

\noi
We observe that we have started with two self-dual $2-$forms
$B, \ttB$ and two anti-self-dual $2-$forms $B', \ttB'$
and we end up with two $2-$forms $B_1, \ttB_1$.
The $2-$forms $B_2, \ttB_2$  are related to the $2-$forms
 $B_1, \ttB_1$ by duality transformations (\ref{B1-B2})
and thus they do not appear in the Lagrangian.
These final $2-$forms are neither self-dual nor anti-self-dual,
which is in perfect agreement with the theory of representations
of the $4D-$Poincar\'e group \footnote{In the representation theory
of $SO(1,3)$, the $2-$forms are either  self- or anti-self-dual. 
For the Poincar\'e group, in the massless case where the little group 
is $SO(2)$, it is  the $1-$forms that are   self or anti-self-dual
(in the case of the electromagnetism these two possibilities  correspond
to the two polarisations of the photon)}. \\

Secondly,  we observe that the terms in the first line of (\ref{free4_2})
are not diagonal. Thus, we define

\beqa
\label{diag}
\hat \varphi_1 = \frac{\varphi_1 + \ttP_1}{\sqrt{2}},
{\hat{ \hat \varphi}}_1 &=& \frac{\varphi_1 - \ttP_1}{\sqrt{2}},
\hat \varphi_2 = \frac{\varphi_2 + \ttP_2}{\sqrt{2}},
{\hat{ \hat \varphi}}_2 = \frac{\varphi_2 - \ttP_2}{\sqrt{2}}, \nonumber \\
\hat B_1 &=& \frac{B_1 + \ttB_1}{\sqrt{2}},
{\hat{ \hat B}}_1 = \frac{B_1 - \ttB_1}{\sqrt{2}},
\eeqa

\noi
 and ${\cal L}_0$ reduces to

\beqa
\label{fermi}
 {\cal L}_0 &=&
\frac12 \partial_m \hat \varphi_1 \partial^m  \hat \varphi_1
- \frac12 \partial_m {\hat {\hat \varphi}}_1 
\partial^m  {\hat {\hat \varphi}}_1 -
\frac12 \partial_m \hat \varphi_2 \partial^m  \hat \varphi_2
+ \frac12 \partial_m {\hat {\hat \varphi}}_2 
\partial^m  {\hat {\hat \varphi}}_2 \nonumber \\
&-&\frac12 \partial_m \tA_1{}_n \partial^m \tA_1{}^n 
+\frac12 \partial_m \tA_2{}_n \partial^m \tA_2{}^n \\
&+&\frac14 \partial_m \hat B_1{}_{np} \partial^m \hat B_1{}^{np}
 -\frac14 \partial_m {\hat {\hat B}}_1{}_{np} \partial^m {\hat {\hat B}}_1{}^{np}.   \nonumber 
\eeqa

\noi
which we now express only in terms of the potentials, including the 
contribution of the gauge fixings terms of (\ref{free4_2}).   
We observe that the kinetic terms for  
${\hat {\hat \varphi}}_1,{\hat { \varphi}}_2,
 \tA_2, {\hat {\hat B}}_1$ have  wrong  relative signs.
We will come back to this point in the next subsection. \\

As it has been noted previously in \cite{cubic}, 
$P^2$ is a Casimir operator, and thus all states
in an irreducible representation have the same mass $m$.
An invariant mass term for each multiplets in (\ref{4-decomposition}) can thus 
be added to the Lagrangian. For instance

\beqa
\label{mass}
{\cal L}[\Xi_{++}]_{\mathrm{mass}}=
m^2(\varphi {\tilde {\tilde \varphi}} +\frac14 B^{mn} \ttB_{mn}
-\frac12 \tA_m \tA^m),
\eeqa

\noi
where the mass $m$ could be related to the parameter $\Lambda$ appearing
in (\ref{matirred}).  
Finally, let us also note that a term like

\beqa
\label{tadpole}
{\cal L}_\varphi= g \ttP
\eeqa

\noi
is invariant on its own. 
We see that this term  is of gradation $1$ which is not the case for
(\ref{free4}), (\ref{mass}) and all the other Lagrangians  considered 
here. 

A last comment regarding the $2$-forms is in order. 
 In (\ref{free4}),  the $2$-forms are self-dual
or  anti-self-dual, so that a usual  gauge transformation
which does not preserve  their (anti)-self-dual character cannot be applied.
The status of gauge fixing through terms of the type
$\frac{1}{12} H_{mnp} {\tilde {\tilde H}}^{mnp} 
-\frac12 {}^\star H_m {}^\star {\tilde {\tilde H}}^m$ in 
(\ref{free4}) is therefore not explicit.
After performing the change(s) of variables (\ref{real}, \ref{diag}), 
the $2$-forms are now neither self-dual nor anti-self-dual, the usual gauge 
transformations become well defined and the gauge fixing for the
$2$-forms in (\ref{free4_2}) (or (\ref{fermi}))
is transparent. We will come back to this point in more details
in section {\bf $4$}.

\subsection{Dualisation}
In this subsection, we propose a possible way to construct
a Lagrangian with correct signs for the various kinetic terms,
based on a special choice for the physical fields. 
The main idea is  related to Hodge duality.
However, the duality transformation will act here on the $p-$forms 
with respect to the Lorentz group $SO(1,3)$. This should be contrasted with
the case of 
the usual duality transformations (generalising the electric-magnetic duality)
which act on the field strengths with respect to  $SO(1,3)$, 
or equivalently on the potentials themselves but with respect
to the little group $SO(2)$. 

To simplify, we use  the notations of
differential forms. Introducing the exterior derivative $d$
which maps a $p-$form into a $(p+1)-$form, and its adjoint $d^\dag$
which maps a  $p-$form into a $(p-1)-$form, we have

\noi
for a $0-$form, say ${\hat {\hat \varphi}}_1$

$$\frac12 \partial_m {\hat {\hat \varphi}}_1\partial^m 
{\hat {\hat \varphi}}_1 = \frac12 d {\hat {\hat \varphi}}_1d {\hat {\hat \varphi}}_1
$$

\noindent
for a $1-$form, say $A_2$ 

$$\frac14 F_2{}_{mn}F_2{}^{mn} + \frac12 \left(\partial_m A_2{}^m\right)^2
=\frac14 d A_2 d A_2 + \frac12 d^\dag  A_2 d^\dag A_2$$

\noi
for a   $2-$form, say ${ {\hat B}}_1$

$$
\frac{1}{12} { {\hat H}}_1{}_{mnp} { {\hat H}}_1{}^{mnp} +
\frac12 \partial^n \hat B_1{}_{nm} \partial_p \hat B_1{}^{pm} 
=\frac{1}{12} d { {\hat B}}_1 d { {\hat B}}_1
+\frac12 d^\dag { {\hat B}}_1 d^\dag { {\hat B}}_1.$$

\noindent
Using,  for a given $p-$form $A_{[p]}$,

\beqa
\label{dd}
&&\frac{1}{(p+1)!} d A_{[p]}  d A_{[p]}  + \frac{1}{(p-1)!} d^\dag A_{[p]}  
d^\dag A_{[p]} 
\\
&=&-\left( \frac{1}{(4-p-1)!} d^\dag  B_{[4-p]}  
d^\dag   B_{[4-p]}  + 
\frac{1}{(4-p+ 1)!} d B_{[4-p]}  d  B_{[4-p]} \right) 
\nonumber 
\eeqa 
\noi
with $B_{[4-p]}= {}^\star A_{[p]}$, and introducing

\beqa
\label{dual}
\begin{array}{ll}
{\hat {\hat D}}_1= {}^\star {\hat {\hat \varphi}}_1,
{\hat { D}}_2= {}^\star  {\hat { \varphi}}_2, 
&\mbox{$4-$forms,}\cr
 \tilde C_2={}^\star \tA_2, & \mbox{$3-$form,} \cr
{\hat {\hat {\cal B}}}_1={}^\star{\hat {\hat B}}_1,
& \mbox{$2-$form,} 
\end{array} 
\eeqa

\noi
we observe that their corresponding kinetic terms
have the correct sign. This means that the physical fields are not
${\hat {\hat \varphi}}_1, {\hat { \varphi}}_2,\tA_2, {\hat {\hat B}}_1$ but
their Hodge duals
${\hat {\hat D}}_1, {\hat { D}}_2, \tilde C_2, {\hat {\hat {\cal B}}}_1$.
The transformation (\ref{dd}) is possible due to the specific form
of our Lagrangian which contains usual kinetic terms plus gauge fixing
terms. In our transformations,  $A_{[p]} \to B_{[4-p]}$,  the kinetic
term of $A$ becomes the gauge fixing term of $B$ and {\it vice-versa}.
This is our duality symmetry. 
In the case of the $0-$form (resp. $4-$form), we only have a kinetic 
(resp. gauge fixing)  term.

At the very end the Lagrangian writes

\beqa
\label{Lagrangien_libre}
\tilde {\cal L}_0&=&
\frac12 d \hat \varphi_1 d \hat \varphi_1  
+\frac12  d {\hat {\hat \varphi}}_2 d {\hat {\hat \varphi}}_2 \nonumber \\
&-& \frac14 d \tA_1 d \tA_1 -\frac12 d^\dag \tA_1 d^\dag \tA_1 \nonumber \\
&+&\frac{1}{12} d {\hat  B}_1 d {\hat  B}_1 + \frac12 d^\dag {\hat B}_1 
d^\dag {\hat B}_1 +
\frac{1}{12} d {\hat {\hat {\cal B}}}_1 d {\hat {\hat {\cal B}}}_1 
+ \frac12 
d^\dag {\hat {\hat {\cal B}}}_1 d^\dag {\hat {\hat {\cal B}}}_1
 \\
&-& \frac{1}{48} d \tilde C_2 d \tilde C_2  - \frac{1}{4} 
d^\dag \tilde C_2 d^\dag \tilde C_2  \nonumber \\
&+&\frac{1}{12} d^\dag {\hat {\hat D}}_1  d^\dag {\hat {\hat D}}_1
+\frac{1}{12} d^\dag { {\hat D}}_2  d^\dag { {\hat D}}_2
\nonumber
\eeqa

\noi
with the physical degrees of freedom as follows: 
in the sector of gradation $-1$ and $1$ two
$0-$forms ($\hat \varphi_1, {\hat {\hat \varphi}}_2$), 
two $2-$forms
(${\hat B}_1, {\hat {\hat {\cal B}}}_1$) and  two $4-$forms 
(${\hat {\hat D}}_1, \hat D_2$); in the zero-graded sector one
$1-$form $\tA_1$ and one $3-$form $\tilde C_2$. 
We note that the physical states are mixtures of states belonging
to two CPT-conjugate multiplets (\ref{real}) and 
also  mixtures of the
graded $(-1)-$ and the graded $1-$sectors (\ref{diag}).\\

Considering the gauge transformation for a $p-$form 
$A_{[p]}$,  ($p \ge 1$),

\beqa
\label{gauge}
A_{[p]} \to A_{[p]} + d \chi_{[p-1]}, 
\eeqa

\noi 
where $\chi_{[p-1]}$ a $(p-1)-$form,
the presence of terms involving $d^\dag$ in the Lagrangian fixes  partially 
the gauge to  
$ d^\dag  d \chi_{[p-1]}=0$. This means that 
the terms involving the 
$d^\dag$ operators
can be seen as some Feynman gauge fixing terms adapted for $p-$forms.
Another way of seeing this phenomenon is to
rewrite ${\cal L}_0$ with  Fermi-like terms 
($\frac{1}{p!} \partial A_{[p]} \partial A_{[p]}$).
It is well known that  $A_{[p]}, 0\le p\le 2$ give rise
to a massless state in the  $p-$order  antisymmetric representation of
the little group $SO(2)$.  But, in our decomposition, there are  also
 $p-$forms with $p = 3,4$. 
[It is interesting to note that
similar phenomena  are well-known  
 in the context of type IIA, IIB string theory
\cite{pol} in $10$ space-time dimensions where $9-$ and $10-$forms appear. 
Actually, subsequent to the early works on  two-forms in
\cite{cs, nambu}, several authors  studied the classical and
quantum properties of the non-propagating $3$- and $4$-forms 
\cite{3f,4f}. In particular,  it was pointed out in \cite{4f} that the gauge
fixing term for a  $4$-form takes the form of a kinetic term for
a scalar field, in exact analogy with our results.]\\

Before ending this section let us make some comments on the number of degrees
of freedom of the various fields and the role played by the gauge fixing
terms [more details are given in section $4$]. We should first stress  the
difference between our case and the conventional gauge invariant theories, 
despite the presence of the ``familiar'' gauge fixing terms of the form
$(d^\dag A_{[p]})^2$ in (\ref{Lagrangien_libre}). Indeed, while in the case of  gauge theories the gauge invariance guarantees that the physical (on-shell) 
quantities are gauge fixing independent, in our case $(d^\dag A_{[p]})^2$
cannot be traded for any other gauge-fixing function since it is imposed by the 3SUSY invariance,  and is thus expected to affect the physical degrees of 
freedom. Let us illustrate the point
on a generic Lagrangian of the form $ L_A = (d A_{[p]})^2+ (d^\dag A_{[p]})^2$ 
which, apart from relative coefficients which are unimportant for the 
discussion, is the one dictated by the 3SUSY invariance for $p=1, 2, 3$. 
$L_A$ has a restricted invariance under
$A_{[p]} \to A_{[p]} + d \chi_{[p-1]}$ and $A_{[p]} \to A_{[p]} + d^\dag \chi_{[p+1]}$ for the subclasses of forms satisfying $d^\dag d \chi_{[p-1]} =0$ and 
$d d^\dag \chi_{[p+1]}= 0$. However due to Poincar\'e's theorem (barring topological effects which we do not consider in this paper) the latter constraint on 
$\chi_{[p+1]}$
implies  that there exists a $(p-1)-$form $\lambda_{[p-1]}$ such that   
$d^\dag \chi_{[p+1]}= d \lambda_{[p-1]}$. Hence the second  invariance of $L_A$
is  actually also of the gauge type with the constraint 
$d^\dag d \lambda_{[p-1]} =0$ and does not correspond to an extra
freedom.\footnote{Note that an equivalent formulation holds if one considers the constraint on $\chi_{[p-1]}$, and amounts to interchanging the roles of
 $d$ and $d^\dag$.}
This shows that the effective degrees of freedom of $A_{[p]}$ are dictated
only by  the gauge freedom eq.(\ref{gauge}), supplemented by
 $d^\dag d \chi_{[p-1]} =0$. An immediate consequence of the latter
constraint is that the usual Lorentz condition $d^\dag A_{[p]}=0$
cannot be imposed in general to eliminate the unphysical components. 
This means that the way one should eliminate the unphysical components cannot
be handled in a usual manner \cite{ghosts}.
On top of that, such a condition is not stable under our transformation laws 
(\ref{transfo2}). [For instance,
if we put $\partial_m \tA^m=0$, then $\partial_m \delta_\e \tA^m=0$ gives 
$\e^m \partial_m \ttP=0$, which is obviously too strong.] 
Furthermore, it should be clear that the constraint $d^\dag d \chi_{[p-1]} =0$
on the gauge generators  $\chi_{[p-1]}$ is not be confused with
the dependences among the gauge transformations on $A_{[p]}$ which originate 
when these transformations possess themselves some gauge invariance  
(see for instance \cite{gomisetal} for a review). Off-shell, the usual gauge invariance would 
have lead to $3$ degrees of freedom for each $1-$ and $2-forms$, and $1$
degree of freedom for the $3-$form. Recall that this conventional counting 
corresponds to the maximal elimination of gauge redundancies {\sl assuming that
 the field components have arbitrarily general forms}. (for instance, particular
configurations such as fields of the pure gauge form can be completely gauged
away.) This point is of particular relevance to our case: the equation
$d^\dag d \chi_{[p-1]} =0$ reduces the space of allowed space-time configurations of $\chi_{[p-1]}$. The elimination of redundant degrees of freedom in
$ A_{[p]}$ are thus possible only when the space-time configurations of
the latter are consistent with those of $\chi_{[p-1]}$. Thus if we insist
on having arbitrary configurations for the components of $ A_{[p]}$, then the 
residual gauge invariance does not eliminate any degree of freedom, {\sl i.e.}
$\tilde A$, $\tilde C$ have each $4$ degrees of freedom, 
${\hat {\hat {\cal B}}}$ and ${\hat  B}$ $6$ degrees of freedom each
and $\varphi$'s and $D$'s one degree of freedom. However the situation
is not as simple, since on the one hand the 3SUSY could itself impose some 
space-time configuration constraints on the components of a given 3SUSY 
multiplet, and on the other hand the gauge transformation should preserve
the 3SUSY character of the transformed fields. We will come back to these 
issues in more detail in section $4$.

\subsection{Mixing between different multiplets}

We  can now consider coupling  terms between different multiplets. 
The basic idea is to couple different types of multiplets 
such that  zero-graded couplings between a potential and a field
strength are possible. Having this
in mind, one can {\it a priori} couple the multiplet
$\Xi_{++}$ with either $\Xi_{+-}$ or $\Xi_{-+}$ (\ref{4-decomposition}).
Imposing the 3SUSY invariance,  $\Xi_{++}$ can only  be coupled
with $\Xi_{+-}$ (or $\Xi_{--}$ with $\Xi_{-+}$). 
We name these two pairs of multiplets 
{\it interlaced multiplets}.

For the sake of simplicity, 
we do not consider hereafter the fields introduced in  the previous
subsection, keeping in mind that the dualisation  (\ref{dual})
can be performed at any step if necessary.

The simplest Lagrangian mixing
$\Xi_{++}$ with $ \Xi_{+-}$  and  
$\Xi_{--}$ with $ \Xi_{-+}$,
expressed with the fields appearing in (\ref{4-decomposition}), is

\beqa
\label{coup}
&{\cal L}_{\mathrm{c}}& = {\cal L}_{\mathrm{c}}(\Xi_{++},\Xi_{+-}) + {\cal L}_{\mathrm{c}}(\Xi_{--},\Xi_{-+})
\nonumber \\
&=&\lambda \left(
\partial_m \varphi \ttA^\prime{}^m 
+\partial_m \ttP  A^\prime{}^m
-\partial_m \tA^m \tP^\prime - \partial_m \tA_n \tB^\prime{}^{mn}
+\partial^m  B_{mn}  \ttA^\prime{}^n
 +\partial^m \ttB_{mn} A^\prime{}^n
\right)  \nonumber \\
&+&
\lambda^\star \left(
\partial_m \varphi' \ttA{}^m 
+\partial_m \ttP'  A^m
-\partial_m \tA'^m \tP - \partial_m \tA'_n \tB^{mn}
+\partial^m  B'_{mn}  \ttA^n
 +\partial^m \ttB'_{mn} A^n
\right)  \nonumber \\
\eeqa

\noi
with $\lambda=\lambda_1 + i \lambda_2$ a complex  coupling constant
with mass dimension.
Due to the CPT conjugation, the Lagrangian ${\cal L}_c$ is real.
We emphasize here that if
terms of non-zero gradation  were included, 
they would have had to be separately invariant as can be seen
from  (\ref{transfo2}).  Furthermore, one can explicitly check that there is 
no invariant Lagrangian which is bilinear in the fields and of gradation $1$ 
or $(-1)$. This is in  perfect agreement with the results of the next section.\\

\medskip
To show the invariance of ${\cal L}_{\mathrm{c}}$ it is sufficient to study 
separately the two parts  $ {\cal L}_{\mathrm{c}}(\Xi_{++},\Xi_{+-})$ and  
$ {\cal L}_{\mathrm{c}}(\Xi_{--},\Xi_{-+})$, since they do not mix under 
3SUSY transformations. From (\ref{transfo2}) one finds

\beqa
\label{var}
 \delta_\varepsilon {\cal L}_{\mathrm{c}}(\Xi_{++},\Xi_{+-}) \hat =
-\frac14 \lambda \tB'^{mn}
 \left(\e^r \partial_{[m} \ttB_{n]_-r} -\e_{[m} \partial^r \ttB_{n]_-r}\right)
\eeqa

\noindent
where we used the shorthand notation $ X_{[m n]_-}$ for the 
anti-self-dualised $2-$form (see eq.(\ref{dualisation})), (the hatted equality
denotes equality up to surface terms).
After some algebraic manipulations (see section {\bf 3.1}) 
one finds 

\beqa
\e^r \partial_{[m} \ttB_{n]_-r} -\e_{[m} \partial^r \ttB_{n]_-r}=0,
\nonumber
\eeqa

\noindent
so that $\delta_\varepsilon {\cal L}_{\mathrm{c}}(\Xi_{++},\Xi_{+-})$
reduces to a total derivative. 
A similar result holds for
${\cal L}_{\mathrm{c}}(\Xi_{--},\Xi_{-+})$, thus (\ref{coup}) is an invariant
Lagrangian.

Let us now consider the total Lagrangian 

\beqa
\label{free}
{\cal L}= {\cal L}_0 + {\cal L}'_0 + {\cal L}_c,
\eeqa

\noi
where ${\cal L}_0$ and ${\cal L}'_0$ are given in (\ref{free4}) and 
${\cal L}_c$ given in (\ref{coup}).
Since ${\cal L}$ is quadratic in the fields, it is always possible, 
by a field
redefinition, to rewrite ${\cal L}$ in a diagonal form. 

To proceed with this calculation, we perform the change of variables defined
in (\ref{real}) and (\ref{diag}) to cast ${\cal L}_0$ in a diagonal form,
(and of course similar transformations to ${\cal L}'_0$). 
Direct inspection shows that
  ${\cal L}$  contains  $15$ fields: 

$6$ scalar fields,   $\hP_1,\hhP_1, \hP_2,\hhP_2$ 
(in ${\cal L}_0$), $\tP_1,\ttP_2$ (in ${\cal L}'_0$);

$6$ vector fields,   $\tA_1,\ttA_2$ (in ${\cal L}_0$), 
$\hA_1,\hhA_1, \hA_2,\hhA_2$ (in ${\cal L}'_0$);

$3$ two-forms        
$\hB_1, \hhB_1$ (in ${\cal L}_0$) $\tB_2$ (in ${\cal L}'_0$).

\noi
The notations for the fields of ${\cal L}'_0$ follow the same logic as
the notations of ${\cal L}_0$. In order to diagonalise ${\cal L}$ one observes that
we have three decoupled Lagrangians: 

\beqa
\label{L123}
{\cal L}= {\cal L}_1(\hP_1,\hP_2,\hA_1,\hA_2,\hB_1)
+{\cal L}_2(\hhP_1,\hhP_2, \hhA_1,\hhA_2,\hhB_1)+{\cal L}_3(\tP_1,\ttP_2, 
\tA_1,\ttA_2,\tB_1).
\eeqa

\noi
with ${\cal L}_1$,${\cal L}_2$ and ${\cal L}_3$ having the same form. 
We will explicitly consider
one of them that we denote generically ${\cal L}(\varphi_1, \varphi_2, A_1, A_2, B)$:

\beqa
\label{L1/3}
{\cal L}(\varphi_1, \varphi_2, A_1, A_2, B)&=&
\frac12 (\partial_m \varphi_1)^2- \frac12 (\partial_m \varphi_2)^2- \frac12 
(\partial_m A_1{}_n)^2 +\frac12 (\partial_m A_2{}_n)^2
+\frac14 (\partial_m B_{np})^2 \nonumber \\
&+&\lambda_1\left(A_1{}^m \partial_m \varphi_1 + A_2{}^m \partial_m \varphi_2
 - B^{mn} \partial_m A_1{}_n - {}^\star B^{mn} \partial_m A_2{}_n
\right)\\
&+&\lambda_2\left(-A_2{}^m \partial_m \varphi_1 + A_1{}^m \partial_m \varphi_2
 + B^{mn} \partial_m A_2{}_n - {}^\star B^{mn} \partial_m A_1{}_n
\right). \nonumber 
\eeqa

\noi
Let us comment at this level some of the terms appearing in this Lagrangian. 
The last two lines of (\ref{L1/3}), which originate from
from (\ref{coup}), contain  exactly the same gauge fixing
as in (\ref{fermi}).  For  $B$,  only 
$\frac12 B^{mn} F_i{}_{mn}$ ($i=1,2$) fix the gauge, 
while ${}^\star B^{mn} F_i{}_{mn}, \ i=1,2$
are  gauge invariant. Terms like  ${}^\star B^{mn} F_{mn}$
(called $BF-$terms) related to topological theories where initially introduced
 in \cite{topo}. Their natural appearance within 3SUSY
may suggest some underlying topological properties whose study is,
however, out of the track of the present paper.  
Other mixing terms between the $A$ and $\varphi$ fields are of the Goldstone
type which appear after spontaneous symmetry breaking, however, in our case
such terms cannot be gauged away since the gauge is already partially fixed.


In order to diagonalise (\ref{L1/3}) we proceed in several steps. 
(Of course in addition to (\ref{L1/3}) one can add the mass terms
(\ref{mass}). But for simplicity we do not consider them here.)
Firstly, we express the action in  
Fourier space. Secondly, we  construct some  perfect squares for the
terms involving   $A_1$ and then   $A_2$. After a tedious calculation, we 
obtain

\beqa
\label{L-diag}
\tilde {\cal L} &=&\frac12\left(p^2-(\lambda_2^2-\lambda_1^2)\right)
 \tP_1(p)\tP_1(-p) 
               - \frac12\left(p^2-(\lambda_2^2-\lambda_1^2)\right) 
\tP_2(p)\tP_2(-p)   \nonumber \\
&+& \lambda_1 \lambda_2   \left( \tP_1(p) \tP_2(-p) + \tP_2(p) \tP_1(-p)
 \right) \nonumber \\
&-& \frac12 p^2 \tA'_1{}_m(p)  \tA'_1{}^m(-p)    + \frac12 p^2 \tA'_2{}_m(p) 
 \tA'_2{}^m(-p) +
\frac14    p^2 \tB_{mn}(p)  \tB^{mn}m(-p)  \\
&+&\frac12 \frac{1}{p^2} p^r p_s(\lambda_1^2 -\lambda_2^2)\left(
\tB_{rm}(p) \tB^{sm}(-p)- {}^\star \tB_{rm}(p) {}^\star \tB^{sm}(-p)\right)
 \nonumber \\
&+&\frac{\lambda_1 \lambda_2}{p^2}  p^r p_s
\left(\tB_{rm}(p) {}^\star \tB^{sm}(-p)+ {} \tB^{s m}(p) {}^\star 
\tB_{rm}(-p)\right), \nonumber
\eeqa

\noi
where the tilde denotes the Fourier transform (not to be confused with the tilde in
the fields we had until  (\ref{L123})) and 

\beqa
\tA'_1{}_m(p)= \tA_1{}_m(p)  +
 \frac{ \lambda_1}{p^2}i p_m \tP_1(p) + \frac{  \lambda_2}{p^2}i p_m \tP_2(p)
+  \frac{ \lambda_1}{p^2}i p^r\tB_{rm}(p) + \frac{ \lambda_2}{p^2}ip^r
({}^\star \tB_{rm}(p))  \\
\tA'_2{}_m(p)= \tA_2{}_m(p)  - \frac{ \lambda_1}{p^2}ip_m \tP_2(p) + 
\frac{ \lambda_2}{p^2}ip_m \tP_1(p)
+  \frac{ \lambda_2}{p^2}ip^r \tB_{rm}(p) - \frac{ \lambda_1}{p^2}ip^r
 ({}^\star \tB_{rm}(p)). \nonumber 
\eeqa

A simple transformation diagonalises the $\varphi$ part  of the Lagrangian. Then, in 
order not
to have unwanted tachyons  we need to impose $\lambda_2^2 \ge \lambda_1^2$. 
 Finally, 
the kinetic part
for the $B$ field is non conventional. Nevertheless,  taking
 $\lambda_1 =0$ simplifies somewhat the Lagrangian.
As expected, the wrong signs  of some of the kinetic terms do not change by 
this diagonalisation.
However, we could have proceeded along the same lines with the fields given in 
section 2.2 where the Lagrangian involves one $0-$, one $1-$, one 
$2-$, one $3-$ and one $4-$form fields.

\section{Interactions}
\renewcommand{\theequation}{3.\arabic{equation}}   
\setcounter{equation}{0}
In the previous section only quadratic terms describing freely propagating
fields (albeit with some non-trivial mixing) were considered. 
To construct interactions one must consider higher 
order  terms.  We will show in this section that 
such terms, describing interactions among the four multiplets 
$\Xi$, are forbidden by 3SUSY. Such an obstruction is welcome, at least
at the level of dimension four operators,
in order to maintain the physical interpretation of the compatibility between 
3SUSY and the $U(1)$ gauge symmetry in terms of gauge fixing of the latter. 
Indeed, such an interpretation would be lost if the 3SUSY allowed
dimension four self-interactions between the gauge fields, which would 
then break further the $U(1)$ gauge symmetry.

\subsection{Derivative multiplets}
In this subsection we state some useful properties which will be repeatedly
used in the rest of the paper\\

\noindent
\underline{{\sl Self-dualities}:}
 denoting generic (anti)-self-dual 2-forms by ($R^{(-)}$) $R^{(+)}$, one has

\begin{equation} \label{slfdual}
\frac12 \varepsilon_{m n p q} R^{(\pm) p q} = \pm i R^{(\pm)}_{m n}
\end{equation}

\noindent 
leading to the following relations

\beqa
R^{(\mp)}_{p q} R^{(\pm)}{}^{p q} &=& 0 \nonumber \\
R^{(\pm)}_{m r} R^{(\pm)}{}_n{}^r &=& \frac14 \eta_{m n} R^{(\pm)}_{p q} R^{(\pm)}{}^{p q}  \label{dualprop1} \\
\varepsilon_{m n p r} R^{(\pm)}{}_q{}^r &=&  \pm i (\, \eta_{m q} R^{(\pm)}_{n p} +\eta_{n q} R^{(\pm)}_{p m} + 
                                   \eta_{p q} R^{(\pm)}_{m n} \,). \nonumber 
\eeqa

\noindent
Furthermore, defining partial derivatives with respect to $R^{(\pm)}_{m r} x^r$, 

$$\bar \partial^m_{(\pm)} \equiv \frac{\partial^m}{\partial R^{(\pm)}_{m r} x^r}$$ 

\noindent
one has

\beqa \label{dualprop2}
\bar \partial^m_{(\pm)} \!\!\! &=& \frac{4}{  R^{(\pm)}_{p q} \; R^{(\pm)}{}^{p q} }R^{(\pm)}{}^m{}_r \; \partial^r \nonumber \\
\overline{ \Box}_{(\pm)} &=& \frac{4}{R^{(\pm)}_{p q} \; R^{(\pm)}{}^{p q}} \; \Box 
\eeqa

\noindent
as a consequence of the second equation in (\ref{dualprop1}) (provided that
$R^{(\pm)}_{p q} \; R^{(\pm)}{}^{p q} \neq 0$). Finally, the 
third equation in (\ref{dualprop1}) leads immediately to

\beqa \label{dualprop3}
\varepsilon_{m n p r}\,  \partial^q \, R^{(\pm)}{}_q{}^r &=&  \pm i (\,
\partial_m R^{(\pm)}_{n p} +\partial_n R^{(\pm)}_{p m} + 
                                   \partial_p R^{(\pm)}_{m n} \,) \nonumber \\
\varepsilon_{m n p r}\,  \partial^p \, R^{(\pm)}{}_q{}^r &=& \pm i (\, 
\partial_q R^{(\pm)}_{m n} + \eta_{q m} \partial_r R^{(\pm)}{}_n{}^r -
\eta_{q n} \partial_r R^{(\pm)}{}_m{}^r \,).
\eeqa

\noindent
We also denote the (anti)-self-dualisation of any second rank tensor $X_{m n}$
by

\begin{equation} \label{dualisation}
X_{[m n]_\pm} \equiv X_{m n} - X_{n m} \mp i \varepsilon_{mnpq} X^{p q}.
\end{equation} 

\noindent
\underline{{\sl Derivative multiplets:}} For each multiplet of a given type
$\Xi$, (\ref{4-decomposition})  we define its derivative ${\cal D} \, \Xi$ by 
saturating properly the Lorentz indices and combining components respecting 
the $\mathbb Z_3-$gradation as well as the self-duality properties in the 
following way:

\noindent
 for $\Xi_{\pm \mp} \equiv (A_m, \; \tP, \tB_{mn}, \; \ttA_m)$ 
one constructs

\beqa
\label{partialXipm}
{\cal D} \,\Xi_{\pm \mp} &=& \Big(\psi, \psi_{mn}, \; \tpsi_m, \; \ttpsi, \ttpsi_{mn} \Big) \nonumber \\
 &\equiv&  \Big(\partial_m A^m, \partial_{[m}A_{n]_\pm}, \; \partial_m \tP+ 
\partial^n \tB^{(\mp)}_{nm}, \; \partial_m \ttA^m, \partial_{[m}\ttA_{n]_\pm}\Big). 
\eeqa

\noindent
Similarly, for  $\Xi_{\pm \pm} \equiv (\varphi, B_{mn}, \;  \tA_m,  \; \ttP, \ttB_{mn})$,

\beqa
\label{partialXipp}
{\cal D} \, \Xi_{\pm \pm} &=& \Big( \psi_m, \; \tpsi, \tpsi_{m n}, \; \ttpsi_m \Big) \nonumber \\
 &\equiv& \Big( \partial_m \varphi + \partial^n B^{(\pm)}_{nm}, \; \partial_m \tA^m, \partial_{[m}\tA_{n]_\mp}, \;  \partial_m \ttP+  \partial^n \ttB^{(\pm)}_{nm} \Big)
\eeqa

\noindent
The transformation laws for ${\cal D} \Xi$ are rather straightforward
to establish using (\ref{transfo2}). For instance, one obtains for
${\cal D} \Xi_{\pm \mp}$

\beqa
&&\delta_\e \psi= \e^m \tpsi_m, \ \
\delta_\e \ttpsi= \e^m \partial_m \psi, \ \ 
\delta_\e \tpsi^m = \e^m \ttpsi + \e_n \ttpsi^{mn} \nonumber \\
&&\delta_\e \psi_{mn}= \e_n \tpsi_m - \e_m \tpsi_n + i \e_{mnpq} e^p \tpsi^q 
+ (\e^r \partial_{[n} \tB^{(\mp)}_{m]_\pm r} - \e_{[n}\partial^r \tB^{(\mp)}_{m]_\pm r} ) \\
&&\delta_\e \ttpsi_{mn}= \e^r \partial_r \psi_{mn}  
 \nonumber 
\eeqa

\noindent
showing that ${\cal D} \Xi_{(\pm \mp)}$ transforms like a $(\pm, \pm)$ 
multiplet provided that

\begin{equation} \label{dual1}
\e^r \partial_{[n} \tB^{(\mp)}_{m]_\pm r} - \e_{[n}\partial^r \tB^{(\mp)}_{m]_\pm r} =0.
\end{equation}

\noindent
This last equation is indeed satisfied as can be shown by using the two
relations in (\ref{dualprop3}). A similar result holds for the transformation
laws of ${\cal D} \, \Xi_{\pm \pm}$. We thus have the following important
property, 

\begin{enumerate}
\item[]{\bf I:} {\it The derivative of any multiplet of the type
$(s \mp)$ is a  multiplet of the type $(s \pm)$.}
\end{enumerate}

\begin{equation}\label{derivX}
{\cal D} \, \Xi_{s \mp} \sim \Xi'_{s \pm}
\end{equation}
(where $s= +$ or $-$).


\bigskip \bigskip
The two next subsections will be devoted to the  proof  that cubic 
supersymmetry
forbids  any interaction terms for the considered  multiplets.

\subsection{Tensor calculus}
A natural way to build  invariant interaction Lagrangians would be to define a 
tensor calculus which allows to construct  a 3SUSY multiplet
starting from two or more multiplets $\Xi$ of any of the four types 
defined in (\ref{4-decomposition})\footnote{Recall that such techniques
were initially used in the case of supersymmetry  
before the superspace formulation \cite{sohnius}.}.
We will show here that if one starts from two arbitrary multiplets 
$\Xi$, it will not be possible to construct quadratically 
a third one of any of the four types defined in (\ref{4-decomposition}). 
In such a systematic study, one has to consider all possible triplets of 
multiplets. We exemplify this here on the specific case  of multiplets
of the type $\Xi_{++}$: starting from
$\Xi_1{}_{++}=\big(\varphi_1, B_1,\tA_1, \ttP_1,\ttB_1\big)$
and $\Xi_2{}_{++}=\big(\varphi_2, B_2,\tA_2, \ttP_2,\ttB_2\big)$  we
seek for a third one of the same type, 
 $\Xi_{12}{}_{++}=\big(\varphi_{12}, B_{12},\tA_{12}, 
\ttP_{12},\ttB_{12}\big)$.

We begin by constructing a scalar $\ttP_{12}$ and require it to transform
like a total derivative.
Using the results of  section 2.1  we have 

$$\ttP_{12}= \varphi_1 \ttP_2 + \varphi_2 \ttP_1 + 
\frac14  B_1{}^{mn} \ttB_2{}_{mn} + \frac14  B_2{}^{mn} \ttB_1{}_{mn}
-\tA_1{}_m  \tA_2{}^m,$$

\noindent
that transforms like a total derivative. This scalar turns out to 
be the only one we can construct (see next subsection).
The transformation law (\ref{transfo2}) $\delta_\e \ttP_{12} = \e^m 
\partial_m \varphi_{12}$
gives (up to a constant)

$$\varphi_{12} = \varphi_1 \varphi_2 + \frac14 B_1{}_{mn} B_2{}^{mn}.$$

\noi
Using  $\delta_\e \varphi_{12}=\e^m \tA_{12}{}_m$, one gets
$$\tA_{12}{}_m= \tA_1{}_m \varphi_2 + \tA_2{}_m \varphi_1 +
\tA_1{}^n B_{2}{}_{nm} + \tA_2{}^n B_{1}{}_{nm}. $$

\noi
Next, applying the transformations 
(\ref{transfo2}) on the R.H.S.
of the equation above,  one finds

\beqa
\delta_\e \tA_{12}{}_m&=&\e_m\left(\ttP_1 \varphi_2 +\ttP_2 \varphi_1 +
2 \tA_1{}_m \tA_2{}^m\right) \nonumber \\ & +&
\e^n \left(\ttB_2{}_{mn} \varphi_1 +B_2{}_{mn} \ttP_1 +
\ttB_1{}_{mn} \varphi_2 +B_1{}_{mn} \ttP_2 +
\ttB_1{}_{pn} B_2{}^p{}_m+ \ttB_2{}_{pn} B_1{}^p{}_m\right).
\nonumber 
\eeqa

\noi
But, in order to have a 3SUSY multiplet, we should have 
$\delta_\e \tA_{12}{}_m = \e_m \ttP_{12} + \e^n \ttB_{12}{}_{mn}$, 
with $\ttB_{12}{}_{mn}$ a self-dual two-form,
which is clearly impossible. Thus we cannot build in such a way
a third $\Xi_{++}$ multiplet
starting from two $\Xi_{++}$ multiplets.\\

Similar calculations can be done for all possible triplets of
multiplets, leading to the same result.

\bigskip \bigskip
Thus, we see  that the simplest idea does not work. Next, in the
following two subsections,
we will  try to construct, in all generality, invariant interacting terms.
For this, we proceed in two main steps.

Firstly,  we start with a given multiplet of the type $\Xi_{\pm \pm}$.
Then, we find the possible sets of fields $\Psi$ (content and transformation 
laws) which couple to $\Xi_{\pm \pm}$ in an invariant way.

Secondly, having obtained $\Psi$, we  would like to get it as
a function $F(\Xi_{++},\Xi_{--},\Xi_{+-},\Xi_{-+})$. This function 
will be proven to be  linear in the fields. Hence, one can get at most 
quadratic terms,
and therefore, no interactions are possible.

\subsection{Possible couplings of a given multiplet}

In this subsection  we couple  a given 3SUSY multiplet $\Xi_{\pm \pm}$ with a
set of fields $\Psi$. Let this multiplet be of the type
$\Xi_{++}=\big(\varphi, B_+, \tA, \ttP, \ttB_+\big)$ with $\varphi, \ttP$
two scalars, $B_+, \ttB_+$ two self-dual $2-$forms  and $\tA$ a vector. 
The other cases ($\Xi_{--}, \Xi_{+-}, \Xi_{-+}$)  are treated along the
same lines.

The most general possibility of  coupling, in a quadratic way,
  $\Xi_{++}$ with the new fields is 

\beqa
\label{X++coup}
{\cal L}({\Xi_{++}, \Psi})=
\varphi \ttpsi + \ttP \psi + \frac14 B_+{}^{mn} \ttpsi_{mn} +
  \frac14 \ttB_+{}_{mn} \psi^{mn} - \tA_m \tpsi^m
\eeqa

\noi
with $\psi,\ttpsi$ two scalars, $\psi_{mn}, \ttpsi_{mn}$ two self-dual
$2-$forms (since $B_-{}_{mn} B_+^{mn}=0$, if $B_-{}_{mn}$
is anti-self-dual) 
and $\tpsi_m$ a vector. In (\ref{X++coup})   {\it a priori}
some of the $\psi$ fields could be set to zero. 
Also, the $\psi$ fields  can or cannot contain 
derivative terms. 
We treat these cases separately.

\begin{enumerate}
\item[]{\bf II:} {\it If the $\psi$ fields contain no derivative terms and 
(\ref{X++coup}) is invariant, then they  form a  multiplet  
of the type $\Xi_{++}$.}
\end{enumerate}

\noi
After an easy calculation, one gets
\beqa
\label{theo1}
\delta_\e {\cal L}(\Xi_{++},\Psi)&=&
\varphi \Big(\delta_\e \ttpsi -\e^m \partial_m \psi\Big)
+\ttP\Big(\delta_\e \psi - \e_m \tpsi^m\Big)
+\frac14 B_+{}_{mn} \Big(\delta_\e \ttpsi^{mn} - \e^p \partial_p \psi^{mn}\Big) 
\nonumber \\ 
&+&\frac14 \ttB_+{}_{mn} \Big(\delta_\e \psi^{mn}  +
\e^m \tpsi^n -\e^n \tpsi^m - i \e^{mnpq} \e_p \tpsi_q\Big) 
 \\
&-&\tA_m\Big(\delta_\e \tpsi^m- \e_n\ttpsi^{mn}  -\e^m \ttpsi\Big)
+\e^p \partial_p\Big(\varphi \psi +\frac14 B_+{}_{mn} \psi^{mn} \Big).
\nonumber 
\eeqa

\noi
By hypothesis, all the fields appearing in $\delta_\e {\cal L}(\Xi_{++},\Psi)$
do not contain derivative terms. This means that no integration by part can 
be done, thus no more total derivatives can be present.
The invariance of ${\cal L}(\Xi_{++},\Psi)$
subsequently means that  $\delta_\e {\cal L}(\Xi_{++},\Psi)=0$ and the
$\psi$ fields transform as a $\Xi_{++}$ multiplet (see (\ref{transfo2})).
This means that all the fields $\psi$ are  present in (\ref{X++coup}).

\bigskip
In the previous case, we did not consider any derivative terms. 
Now, if  we assume
that the $\psi$ fields contain only  first  derivative terms, their  most 
general form is

\beqa
\label{derivative}
&&\psi= \partial_m \lambda^m,\ \  \ttpsi = \partial_m \ttl^m  \nonumber \\
&&\psi_{mn}= \partial_m \lambda'_n -\partial_n \lambda'_m -
i \varepsilon_{mnpq} \partial^p \lambda'^q   \\
&&\ttpsi_{mn}= \partial_m \ttl'_n -\partial_n \ttl'_m -
i \varepsilon_{mnpq} \partial^p \ttl'^q,  \nonumber \\
&&\tpsi_m = \partial_m \tl + \partial^n \tl_{nm}.\nonumber
\eeqa

\noi
(with $\psi_{mn}, \ttpsi_{mn}$ self-dual $2-$forms)
with $\tl$ a scalar, $\lambda_m, \lambda'_m, \ttl_m, \ttl'_m$ four vectors
and $\lambda_{mn}$ a $2-$form (whose anti-(self-)dual character is not
 specified  at that point). 

\bigskip
\begin{enumerate}
\item[]{\bf III:} {\it If the $\psi$ fields are as in (\ref{derivative})  and 
the Lagrangian (\ref{X++coup})
is invariant, then they  form a    
$\Xi_{++}$ multiplet}.
\end{enumerate}

\bigskip
\noi
As before the variation of (\ref{X++coup}) gives (\ref{theo1}). 
It is more natural  to
obtain the variations of the fields $\psi$ instead of the 
ones of the fields $\lambda$.
Since now we have allowed derivative couplings, some integration by part can be
done leading to total derivatives. This means, in particular,
 that {\it a priori}
one cannot  put  $\delta_\e {\cal L}(\Xi_{++},\Psi)=0$.
Paying attention to this possibility one gets

\beqa
\label{theo2}
&&\delta_\e \psi= \e^m \tpsi_m, \ \
\delta_\e \ttpsi= \e^m \partial_m \psi, \ \ 
\delta_\e \tpsi^m = \e^m \ttpsi + \e_n \ttpsi^{mn} \nonumber \\
&&\delta_\e \psi_{mn}= \e_n \tpsi_m - \e_m \tpsi_n + i \e_{mnpq} e^p \tpsi^q \\
&&\delta_\e \ttpsi_{mn}= \e^r \partial_r \psi_{mn}  
 \nonumber 
\eeqa

\noi
Indeed, as in the previous case, no other total derivative can appear.
For instance,  looking at the variation of $\psi_{mn}$ one could have
$\delta_\e \psi_{mn}= \e_n \tpsi_m - \e_m \tpsi_n + i \e_{mnpq} e^p \tpsi^q
+ X_{mn}$ with $B^{mn} X_{mn} $ being a total derivative. Taking into
account the various possibilities to built such an
$X_{mn}$ from the $\lambda$ fields, it is not difficult to  check
that $X_{mn} =0$. Finally, the transformation laws of
the $\lambda$ fields can now be deduced form (\ref{theo2}).
For instance, one gets

\beqa
\label{transfolambda}
\delta_\e \ttl_m = \e^p\partial_p \lambda_m +
a(\e_m \partial_n \lambda^n - \e_n \partial^n \lambda_m) +
a'(\e_m \partial_n \lambda'^n - \e_n \partial^n \lambda'_m),
\eeqa

\noi
where $a,a'$ are arbitrary constants.
Similarly one can obtain the variations of the other $\lambda$ fields,
but it is not necessary for our purpose. 

Note that the coupling Lagrangian  ${\cal L}_c(\Xi_{++}, \Xi_{+-})$ in 
(\ref{coup}) is a special case of the above study  where $\lambda'_m \equiv \lambda_m, \, \ttl'_m \equiv \ttl_m$ in eq.(\ref{derivative}). In this case,
the  $\psi$'s form the derivative multiplet, (\ref{partialXipm}),
 of the $\lambda$'s.

\bigskip
We treat  now the most general case, when 

\beqa 
\label{field3}
&&\psi=  \rho + \partial_m \lambda^m,\ \  
\ttpsi =  \ttr + \partial_m \ttl^m, \nonumber \\
&&\psi_{mn}= \rho_{mn}+ \partial_m \lambda'_n -\partial_n \lambda'_m -
i \varepsilon_{mnpq} \partial^p \lambda'^q,   \\
&&\ttpsi_{mn}= \ttr_{mn}+ \partial_m \ttl'_n -\partial_n \ttl'_m -
i \varepsilon_{mnpq} \partial^p \ttl'^q,  \nonumber \\
&&\tpsi_m = \tr_m+ \partial_m \tl + \partial^n \tl_{nm}.\nonumber
\eeqa

\bigskip
\begin{enumerate}
\item[]{\bf IV:} {\it If the $\psi$ are as in (\ref{field3})  and 
the Lagrangian (\ref{X++coup})
is invariant, then they  transform as in (\ref{theo2})}.
\end{enumerate}

\bigskip
The proof is analogous to the case {\bf III}. Here again the $\psi$ fields 
must all be present in (\ref{X++coup}), 
but now some of the $\lambda$ or $\rho$ fields
can be absent in (\ref{field3}).

In {\bf II}, {\bf III} or {\bf IV}
 we have assumed that the fields $\psi$ contain at
most one derivative. One should of course address the more general case
where higher number of derivatives are allowed. In fact, in
this case also, the results remain unchanged.\\

If one considers terms with two derivatives, the only
scalar, vector and $2$-form that can built starting
from  a scalar  $\lambda$,  a vector $\lambda_m$ 
and a $2$-form $\lambda_{mn}$,   are $\psi = \Box \lambda 
+ \partial^m \lambda_m$,
$\psi_m= \Box \psi_m + a \partial_m \partial^n \psi_n$ and 
$\psi_{mn}= \Box \lambda_{mn} + b \partial_{[m}
 \partial^p \lambda_{n]_+p}$. After introducing the
fields $\Psi=(\psi, \psi_{mn},  \tpsi_m,\ttpsi, \ttpsi_{mn})$, with
two derivatives as above,
the invariance of   (\ref{X++coup}) requires, as 
in the proof  of property {\bf II},  
that $\Psi$ is a $\Xi_{++}$ multiplet. 
When the fields are expressed with pure d'Alembertian, we obtain
nothing else but the  Lagrangian of the type ${\cal L}_0$ in
(\ref{free4}).
If we reiterate the process with an even  number of derivatives, 
similar arguments lead to the same conclusion that property {\bf II}
holds.
For the case of  an odd number of derivatives, property {\bf III} holds.
The two cases are different, because of the possibility of using
identity (\ref{dual1})    for an odd number of derivatives.

As previously stated similar results hold when one studies 
the $\Xi_{--}, \Xi_{+-}$ and $\Xi_{-+}$  multiplets. 

\bigskip
In this subsection, we have shown that the invariance of (\ref{X++coup}) 
implies
some specific behaviour of the fields $\psi$. The next step is to 
construct these $\psi$ fields from the multiplets
$\Xi_{\pm \pm}$.

\subsection{Generalised tensor calculus}
The purpose of this subsection, is to explicitly get, out 
of all the 3SUSY multiplets, the fields $\psi$ found in the
previous subsection. In other words, we want to find  functions

\beqa
\label{F}
F(\Xi_{++}, \Xi_{--},\Xi_{+-},\Xi_{-+})=\Big(
f, f_{mn}, \tf_m,\ttf,\ttf_{mn} \Big), 
\eeqa

\noi
where  $f, f_{mn}, \tf_m,\ttf,\ttf_{mn}$
depend  on the four multiplets, and transform like in the previous
subsection.  Of course when one writes that a function depends on a
multiplet $\Xi_{\pm \pm}$ this means that it depends on its fields
as well as on  the derivatives of its fields.

In this subsection also we concentrate  on the $\Xi_{++}$ multiplet, 
since the other cases are similar.

\bigskip
{\it {\bf V:} The only function $F$  defined as in (\ref{F}),
with at most first order derivatives in the fields and  transforming as
a $\Xi_{++}$ multiplet is

$$F(\Xi_{++}, \Xi_{--},\Xi_{+-},\Xi_{-+}) = \alpha \Xi_{++} + 
\beta {\cal D} \Xi_{+-},
\alpha, \beta  \in \mathbb C.$$
}

\bigskip
We first consider the case where  no derivative dependence in
the fields is present in the functions $f$. Furthermore, 
 we assume in a first time that $F$ depends only on the first 
multiplet $\Xi_{++}$.
We start by  writing  the transformation laws of the $f$ fields,
like {\it e.g.} 

\beqa
\label{F1}
\delta_\e \ttf = \e^m \partial_m f =
\frac{\partial \ttf}{\partial \ttP} \delta_\e \ttP +
\frac{\partial \ttf}{\partial \varphi} \delta_\e \varphi +
\frac12 \frac{\partial \ttf}{\partial B_{mn}} \delta_\e B_{mn}+
\frac12 \frac{\partial \ttf}{\partial \ttB_{mn}} \delta_\e \ttB_{mn} +
\frac{\partial \ttf}{\partial \tA_{m}} \delta_\e \tA_{m}.
\eeqa

\noindent
Substituting in (\ref{F1}) the variations (\ref{transfo2}) of the fields
we get
\beqa
 \e^m \partial_m f &=&
\frac{\partial \ttf}{\partial \ttP}  \e^m \partial_m \varphi +
\frac{\partial \ttf}{\partial \varphi}  \e^m \tA_m 
-2  \frac{\partial \ttf}{\partial B_{mn}} \e_m \tA_n 
\nonumber \\
&+&
\frac12 \frac{\partial \ttf}{\partial \ttB_{mn}}  \e^p \partial_p B_{mn}+
\frac{\partial \ttf}{\partial \tA_{m}} (\e_m \ttP + \e^n \ttB_{mn}).
\eeqa

\noindent
In the L.H.S. of the equation above we have only one derivative.
In the R.H.S. since $\ttf$ depends only on the fields and not on its
derivatives, we have 

$$
\frac{\partial \ttf}{\partial \varphi}=0,
 \frac{\partial \ttf}{\partial B_{mn}} =0,
\frac{\partial \ttf}{\partial \tA_{m}} =0.
$$

\noindent
Then, after integration by parts (\ref{F1}) reduces to 

$$\partial_m( \ttf -
\frac{\partial \ttf}{\partial \ttP} \varphi  
 - \frac12 \frac{\partial \ttf}{\partial \ttB_{mn}} B_{mn} )
= - \varphi \partial_m 
\frac{\partial \ttf}{\partial \ttP}
-\frac12 B_{np} \partial_m  \frac{\partial \ttf}{\partial \ttB_{np}}.$$

\noindent
Arguing as before, one gets  $\partial_m 
\frac{\partial \ttf}{\partial \ttP}=0, 
 \partial_m  \frac{\partial \ttf}{\partial \ttB_{np}}=0 $, and

$$\ttf= \alpha \ttP + \frac12 X_{mn} \ttB^{mn}.$$

\noindent
The transformation law of $\ttf$ easily gives
$f= \alpha \varphi + \frac12 X_{mn} B^{mn}$. Then, the transformation
law of $f$ gives $\tf_m= \alpha \tA_m - 2 X_{mn} \tA^n$. Finally, the
transformation
law of $\tf_m$ gives $X_{mn}=0$, and $\ttf_{mn}= \alpha \ttB_{mn}$.
Finally we get  

$$F(\Xi_{++}) = \alpha \Xi_{++}.$$

Now, when one takes into account that $F$ depends on all the
multiplets, $F(\Xi_{++}, \Xi_{+-}, \Xi_{+-},$ $ \Xi_{--})$,
one obtains, by arguments along the same lines 
$$F(\Xi_{++}, \Xi_{+-}, \Xi_{+-},\Xi_{--}) = \alpha \Xi_{++}.$$

\noi

Now we treat the general case, namely when $F$ depends on the 
all the multiplets
 (\ref{4-decomposition})
 as well as on 
their first order  derivatives.
In this case the proof is more intricate
since  terms like $\frac{\partial f}{\partial \partial_m \phi}
\delta_\e \partial_m \varphi$  are present
and 
\beqa
\label{lemme4}
F(\Xi_{++}, \Xi_{+-}, \Xi_{+-},\Xi_{--}) = \alpha \Xi_{++} + 
\beta {\cal D} \Xi_{+-}.
\eeqa

\noindent
with  ${\cal D} \Xi_{+-}$ being defined in (\ref{partialXipm}).\\

So far we have considered the possibility to build only 
$\Xi_{++}$ multiplets. One could address the possibility to 
obtain the fields $\lambda$ transforming as (\ref{transfolambda}).
Assume now, that one can  non-linearly  built such $\lambda$
which we denote  generically by $\Lambda = G(\Xi)$. 
From (\ref{derivative}) written generically as $\Psi= \partial \Lambda$,
one can non-linearly obtain $\Psi = \partial G(\Xi)= F(\Xi, {\cal D} \Xi)$.
Since the  $\psi$ fields  $\Psi$ form a $\Xi_{++}$ multiplet  
(see  (\ref{theo2})) such a non-linear function does not exists
(see {\bf V}). This means that there is no non-linear 
functions leading to the $\lambda$ fields.

\bigskip
  The case of functions involving higher number of derivatives goes along the 
same lines. This leads to the new possibilities
$F(\Xi_{++}, \Xi_{+-}, \Xi_{+-},\Xi_{--}) = \alpha \Box^n \Xi_{++}$
(resp. $F(\Xi_{++}, \Xi_{+-},$ $ \Xi_{+-},\Xi_{--}) = 
\alpha \Box^n {\cal D} \Xi_{+-}$
for  an even (resp. odd) numbers of derivatives.

\section{ Compatibility with $U(1)$ gauge symmetry }
\renewcommand{\theequation}{4.\arabic{equation}}   
\setcounter{equation}{0}
In this section we address the question of compatibility between
the 3SUSY and the $U(1)$ gauge transformations. We note first
that the Lagrangians given in Eqs. (\ref{free4}), (\ref{coup}) 
are invariant  under the following 
transformations 

\beqa \label{g-trans}
\phi &\to& \phi + k,  \nonumber \\
{\cal A}_m &\to& {\cal A}_m + \partial_m \chi,   \\
{\cal B}^{(\pm)}_{m n} &\to& {\cal B}^{(\pm)}_{m n} + 
\partial_m \chi_n - \partial_n \chi_m  \mp i \varepsilon_{mnpq} \partial^p \chi^q\nonumber
\eeqa

\noindent
with

\begin{equation} \label{g-fix}
\partial_m k = 0, \;\; \Box \chi = 0, \;\; \Box \chi_n - \partial_n  \partial_m \chi^m = 0 
\end{equation}

\noindent
 where $\phi(x), {\cal A}_{{}_m}(x), {\cal B}^{(\pm)}_{{}_{m n}}(x)$ denote 
generically the $0-$, $1-$ and $2-$form fields appearing in the various 
$\Xi$ multiplets and $(\pm)$ indicates the self-duality properties
 (\ref{slfdual}). The transformations in (\ref{g-trans}) can be qualified
as gauge transformations. Indeed, strictly speaking,
the gauge transformations should be required for the
{\sl real-valued} fields defined in (\ref{real}). However, for the $0-$ and 
$1-$forms these have the same form as those in (\ref{g-trans})
by linearity.
The case of the $2-$forms is somewhat different: the usual gauge transformations
of the real-valued fields (\ref{real}) which read
\begin{eqnarray}
B_1{}_{m n} \to B_1{}_{m n} + \partial_m \chi_n - 
\partial_n \chi_m \nonumber \\
\ttB_1{}_{m n} \to \ttB_1{}_{m n} + \partial_m {\tilde {\tilde \chi}}_n - 
\partial_n {\tilde {\tilde \chi}}_m 
\nonumber \\
\end{eqnarray}

\noindent
lead to the transformations (\ref{g-trans}) for ${\cal B}^{(\pm)}$ 
as a consequence of projecting out the (anti)-self-duality content of the
real-valued $2-$forms (see (\ref{real}), (\ref{B1-B2})).   
Nonetheless, for arbitrary $k, \chi$ and  $ \chi_m$, the transformations 
(\ref{g-trans}) do not preserve in general the 3SUSY multiplet structures, 
so that the
 3SUSY invariance of the gauge transformed Lagrangian loses its meaning.
Namely, it seems difficult to put in the same 3SUSY multiplet
 the gauge parameters
($(k, \chi_m, \tilde \chi, {\tilde{ \tilde k}},
{\tilde {\tilde \chi}}_m)$
 for say
the $\Xi_{++}$ multiplet). 
 It
 is thus mandatory, for the sake of consistence, to seek for subclasses of 
gauge transformations of the form $ \delta_{gauge} \Xi = \Lambda$ where $\Xi$ 
and $\Lambda$ are 3SUSY multiplets of the same type. [Recall that
in the case of usual supersymmetry, this is achieved rather transparently
 in terms of superfields in the form $V \to V + \Phi + \Phi^\dag$ \cite{wb}.]
In the present case, not having a superfield formulation at our disposal, 
we will make use of the derivative multiplets defined in section {\bf 3.1}. \\

Before studying further this
point, a general remark is in order here: the Lagrangians 
(\ref{free4}), (\ref{coup}) are also invariant under a general shift 
transformation $\Xi \to \Xi \pm \Theta_0$ provided that $\Xi$ and 
$\Theta_0= (..., \theta_0, ...,\theta^m_0, ... \theta^{m n}_0, ...)$ 
are of the same type and that the components of $\Theta_0$ satisfy conditions
similar to (\ref{g-fix}), $ \partial_m \Theta_0 \equiv (..., \partial_m 
\theta_0, ..., \partial_m \theta^m_0, ...\partial_m \theta^{m n}_0, ...) =0$ 
together with
$\Box \theta^m_0 =0$,  
but where its $1-$ and $2-$form components do 
not necessarily have to be differential {\sl exact} forms as required by a 
gauge transformation\footnote{We stress that the similarity with (\ref{g-fix})
is a consequence of the (anti)-self-duality of $\theta^{m n}_0$. Indeed,
while the field strengths $H_{mnp}$ (and their duals) appearing in 
(\ref{free4}), are {\sl automatically} invariant under the transformation
of ${\cal B}$ in (\ref{g-trans}), their invariance under this new transformation
(where $\theta^{m n}_0$ is (anti)-self-dual),  
requires $\partial_m \theta^{m n}_0=0$ as can be shown by using the first 
identity in (\ref{dualprop3}).}. Furthermore, 
combining this transformation with
a 3SUSY transformation, one finds an invariance under
$\Xi \to \Xi + \delta_\varepsilon \Theta_0$ as a consequence of the following
series of equalities (up to surface terms):

\beqa
&& {\cal L}(\Xi) \, \hat = \, {\cal L}(\Xi \, + \, \delta_{-\varepsilon} \Xi)
\, \hat = \, {\cal L}(\Xi + \, \delta_{-\varepsilon} \Xi + \, \Theta_0) 
\, \hat = \, 
\nonumber \\
&&  {\cal L}(\Xi + \Theta_0 + \delta_{-\varepsilon} (\Xi + \Theta_0) +
\delta_\varepsilon \Theta_0 ) \, \hat = \, {\cal L}(\Xi + \Theta_0 +
\delta_\varepsilon \Theta_0 ) \, \hat = \, {\cal L}(\Xi + 
\delta_\varepsilon \Theta_0 ). \nonumber
\eeqa

\noindent
It is worth noting that the condition on $\partial_m \Theta_0$ is not
preserved by 3SUSY, as can be seen from the $0-$ and $2-$form transformations
(\ref{transfo2})- that is $ \partial_m \delta_\varepsilon \Theta_0 \neq 0$. 
Thus, the transformation 

\begin{equation} \label{newsym}
\delta \Xi \equiv \delta_\varepsilon \Theta_0
\end{equation}

\noindent
identified above provides indeed a new symmetry.

Let us now consider the specific case of gauge transformations. Since we
need simultaneously derivatives and definite 3SUSY multiplet structures,
one can make use of eq.(\ref{derivX}) and seek for
a transformation of the form

\begin{equation}
\Xi_{s \pm} \to \Xi_{s \pm} + {\cal D} \Lambda_{s \mp}\;\;, \;\;\;\;\;\;\;\;\;\; (s = +, -)
\end{equation}

\noindent
For instance starting from 
$\Xi_{++}=\big(\varphi, B_{m n}, \tA_m, \ttP, \ttB_{m n} \big)$
and $ \Lambda_{+-}= \big(\lambda_m, \tl, \tl_{mn}, \ttl_m \big)$ one has,
(\ref{partialXipm}),

\beqa  \label{g-trans1}
 \varphi &\to& \varphi + \partial_m \lambda^m  \nonumber \\
 B_{m n} &\to&  B_{m n} + \partial_m \lambda_n - \partial_n \lambda_m 
- i \varepsilon_{mnpq} \partial^p \lambda^q \nonumber \\
 \tA_m &\to& \tA_m + \partial_m \tl + \partial^n \tl_{nm} \\
 \ttP &\to& \ttP + \partial^m \ttl_m \nonumber \\
  \ttB_{m n} &\to& \ttB_{m n} + \partial_m \ttl_n - \partial_n \ttl_m 
- i \varepsilon_{mnpq} \partial^p \ttl^q\nonumber
\eeqa

\noindent
The invariance conditions (\ref{g-fix}) read here 
$\partial_m {\cal D} \Lambda =0$ and reduce to (see also footnote 4), 

\beqa
&& \partial_m \; (\partial \cdot \lambda) = \partial_m \; 
(\partial \cdot \ttl) = 0  \label{g-conditions1}\\
&& \Box \lambda_m = \Box \ttl_m =0
 \label{g-conditions2}\\
&& \Box \tl + \partial^m \partial^n \tl_{nm} \equiv \Box \tl = 0 
\label{g-conditions3} 
\eeqa 

%

One further constraint comes from the requirement that the transformation of 
$\tA$ in (\ref{g-trans1}) should be an exact form as in the gauge 
 transformations (\ref{g-trans}),
that is

\begin{equation} \label{g-conditions4}
\partial^n \tl_{nm} \equiv \partial_m \chi
\end{equation} 

\noindent
This constraint is not trivial, and for $\tl_{nm}$ anti-self-dual it implies

\begin{equation} \label{g-conditions4p}
\Box \tl_{nm} =0
\end{equation}

\noindent
as can be proven by using the first equation in (\ref{dualprop3})\footnote{
Obviously, the 
various properties discussed here as well as in section {\bf 3.1} could be
also derived compactly in the language of differential geometry.}. Furthermore,
the anti-symmetry of $\tl_{nm}$ leads trivially to

\begin{equation} \label{g-conditions4pp}
\Box \chi =0.
\end{equation}

\noindent
To proceed, our strategy will be as follows: determine first the general 
functional forms of $\lambda_m, \ttl_m$ satisfying the constraints 
(\ref{g-conditions1}, \ref{g-conditions2}), and those of $\tl, \chi$ 
satisfying (\ref{g-conditions3}, \ref{g-conditions4pp}). Then, knowing $\chi$, 
construct explicitly  a general antisymmetric, anti-self-dual $2-$form
satisfying (\ref{g-conditions4}), which would then automatically satisfy
(\ref{g-conditions4p}). The fact that the components of $\Lambda_{+ -}$
can indeed be {\sl general functions} (in terms of variables yet to be 
identified), despite the gauge fixing conditions 
(\ref{g-conditions1} -\ref{g-conditions3}, 
\ref{g-conditions4p}), is crucial to assess the elimination of unphysical
degrees of freedom of the fields $\Xi$. It is worth stressing that these gauge
fixing conditions do not allow to choose in general Lorentz gauges, $\partial_m \tA^m=0$, or $\partial^n B_{n m}=0 $. Other gauges such
as the Coulomb gauge, the axial gauge, etc... can in principle be imposed. 
However, if for instance the scalar functions 
$\tl$ and $\chi$ depend only on the space-time Lorentz  invariant $x_m x^m$,
then the conditions (\ref{g-conditions3}, \ref{g-conditions4pp}) determine
uniquely their functional form, $\tl[x^2] \sim \chi[x^2] \sim 1/x^2$ up to
some additive constants. In this case only special space-time configurations
of the field $\tA_m$ can be eliminated by a gauge choice. It is thus 
tempting to consider more general trial functions ${\cal F}[x^2, \; 
\xi \cdot x, \; A_{m n} x^m x^n, \cdots]$, where $\xi_m$ and $A_{m n}$ are 
some constant $4-$vector and symmetric tensor. The inclusion of a $4-$vector
$\xi_m$ is somewhat natural in the context of the 3SUSY algebra whose 
generators (and transformation parameters) are also $4-$vectors. On the other 
hand, one can also include a dependence on constant (anti)-self-dual $2-$forms 
$R_{ m n}$ which sit in the same 3SUSY multiplet as $\xi_m$. For instance
 $A_{m n} \equiv R_{m p} R^p{}_n$ induces a dependence on $R^2$ in ${\cal F}$
(see eq.(\ref{dualprop1}) ), while a dependence on invariants such as
$R_{m p} x^m \xi^p$ will turn out to be also natural to consider.\\

To start with, we studied functions of $x^2$ and $\xi \cdot x$ and established
 the following properties.

\begin{enumerate}
\item[]{\bf (i):} {\it  $\Box {\cal F}[x^2, \; \xi \cdot x] = 0$ has generic
solutions if and only if $\xi^2=0$; these solutions take the form:
$${\cal F}(x^2, \; \xi \cdot x) = G[\xi \cdot x] + (\xi \cdot x)^{-1} 
H[\frac{x^2}{(\xi \cdot x)}]$$ where G and H are arbitrary functions.}
\item[]{\bf (ii):} {\it  if $\xi^2 \ne 0$, the general solution for 
$\Box {\cal F}[x^2, \; \xi \cdot x] = 0$ takes the more particular form
$${\cal F}[x^2, \; \xi \cdot x] =((\xi \cdot x)^2 - \frac{\xi^2 x^2}{4} ) (\frac{C_1}{(x^2)^3} + C_2) + 
   C_3 \frac{(\xi \cdot x)}{(x^2)^2} + C_4 \; \xi \cdot x + \frac{C_5}{x^2} + 
C_6$$ 
where the $C_i$ are arbitrary constants.}
\end{enumerate}

\noindent
The above properties determine the general form of the fields $\tl$ and $\chi$
subject to (\ref{g-conditions3}, \ref{g-conditions4pp}).
In the sequel we will stick to the case $\xi^2=0$ since, according
to {\bf (i)}, it allows the most general configurations for the gauge 
transformations. The general form for the fields $\lambda_m, \ttl_m$ is then 
determined by the following property which we established,

\begin{enumerate}
\item[]{\bf (iii):} {\it 
The general solution for a $1-$form ${\cal F}_m[x^p, \xi^p]$
(with $\xi^2=0$), subject to the two constraints, $\Box {\cal F}_m=0$ and
$\partial_m (\partial \cdot {\cal F}) =0 $,  is

$$ {\cal F}_m= g[\xi \cdot x] \xi_m + \alpha x_m + 
(\frac{1}{(x^2)^2} \alpha_{m r} + \beta_{m r}) x^r +
\kappa (\frac{x^2}{(\xi \cdot x)^3} \xi_m - \frac{x_m}{(\xi \cdot x)^2})$$

where $g$ is an arbitrary function, $\kappa, \alpha, \beta_{m n}$ arbitrary
constants and $\alpha_{m n}$ an arbitrary anti-symmetric tensor.}
\end{enumerate}

\noindent
Thus,
\begin{equation} \label{eqtl}
\tl[\xi \cdot x, x^2] = G_1[\xi \cdot x] + 
(\xi \cdot x)^{-1} H_1[\frac{x^2}{(\xi \cdot x)}] 
\end{equation}

\begin{equation} \label{eqchi}
\chi[\xi \cdot x, x^2] = G_2[\xi \cdot x] + 
(\xi \cdot x)^{-1} H_3[\frac{x^2}{(\xi \cdot x)}] 
\end{equation}

\beqa \label{eqlambdam}
\lambda_m [\xi \cdot x,  x^2] &=& g_1[\xi \cdot x] \xi_m + 
 \alpha x_m + (\frac{1}{(x^2)^2} \alpha_{m r} + \beta_{m r}) x^r \nonumber \\
&&
 +\kappa_1 (\frac{x^2}{(\xi \cdot x)^3} \xi_m - \frac{x_m}{(\xi \cdot x)^2}) 
\eeqa

\beqa \label{eqttlm}
\ttl_m [\xi \cdot x, x^2]& =& \tilde{\tilde g}_1[\xi \cdot x]
 \xi_m  + 
\tilde{\tilde \alpha} x_m + (\frac{1}{(x^2)^2} \tilde{\tilde \alpha}_{m r} +
 \tilde{\tilde \beta}_{m r}) x^r \nonumber \\
&& + \tilde{\tilde \kappa}_1 (\frac{x^2}{(\xi \cdot x)^3} \xi_m - 
\frac{x_m}{(\xi \cdot x)^2}) 
\eeqa

\noindent
Starting from (\ref{eqchi}), one can construct explicitly $\tl_{m n}$ 
satisfying the constraint (\ref{g-conditions4}), in the form

\begin{equation} \label{eqtlmn}
\tl_{m n}[\xi \cdot x, x^2] = x_{[m} \xi_{n]_-} F[\xi \cdot x,  x^2] 
\end{equation}

\noindent
The function $F$  can be determined in terms of $G_2, H_3$
appearing in (\ref{eqchi}). 
One finds

\begin{equation}
F[\xi \cdot x, x^2] = - (\xi \cdot x)^{-2} H_3[\frac{x^2}{(\xi \cdot x)}]
+ (\xi \cdot x)^{-1} G_2[\xi \cdot x]
 -2(\xi \cdot x)^{-3} \int \limits_0^{\xi \cdot x} G_2[t]\; t \; dt 
\end{equation}

\noindent

To summarize, we have proven the existence of gauge transformations
which preserve the type of 3SUSY multiplets and satisfy the necessary
constraints for gauge invariance. However, 
the gauge transformation functions are found to be not completely arbitrary. 
As can be seen from eqs.(\ref{eqtl}, \ref{eqchi}) arbitrary gauge fixing can
be  {\sl a priori} applied to arbitrary field configurations of
$\tA_m$ as a function of $\xi \cdot x$, while 
configurations in $x^2$ can be gauge-fixed only in conjunction with
 $\xi \cdot x$. In contrast, gauge transformations
of $B_{m n}$ and $\ttB_{m n}$ correspond only to
specific functions of $x^2$. Indeed,  direct inspection
of (\ref{eqlambdam}, \ref{eqttlm}) shows that $\partial_{[m} \lambda_{n]_-}, (
\partial_{[m} \ttl_{n]_-})$ appearing in (\ref{g-trans1}), receive
contributions only from $\alpha_{m r}, \beta_{m r}, 
(\tilde{\tilde \alpha}_{m r}, \tilde{\tilde \beta}_{m r})$.\\
To eliminate unphysical degrees of freedom for more general field 
configurations, one can make use of the symmetry (\ref{newsym}). However, one 
should keep in mind that even for this non-gauge transformation, the 
required constraints $\partial_m \Theta_0=0$ and $\Box \theta^m_0=0$ will 
somewhat reduce the generality of the field configurations. \\

Hence, it seems that  one is lead, at this stage of the analysis, to the 
unusual feature that the number of physical degrees of freedom of 
the gauge fields depends on the space-time configurations of these fields! 
However a
more thorough study is still needed and is actually akin to the way space-time
itself transforms under 3SUSY. The fact that $x^m$ should transform 
non-trivially under 3SUSY is obvious from the presence of $\partial_m$ in 
(\ref{transfo2}) which implies that the transformation law of a field
of gradation $1$  depends on the space-time 
configuration of its  partner field of gradation $-1$ [e.g. $\ttP$ does not
transform if $\varphi$ is constant in $x^m$, etc...]. A related question, not 
addressed so far, is whether the $0-$, $1-$, and $2-$form fields 
should verify some space-time constraints in order to be members of the
same 3SUSY multiplet. The answer to such a question depends crucially on the 
way  $x^m$ transforms under 3SUSY, which in turn depends on the possibility to 
define a space of parameters including $x^m$ and leading to the correct 
transformations of the fields. We give here for illustration one example of 
how a non-trivial transformation of $x_m$ can induce constraints. Assume that 
$x^m$ belongs to a $(+,-)$ multiplet 
$X_{+ -} = (x_m, \tilde{\alpha}, \tilde{R}_{m n}, \tilde{\tilde{\xi}}_m)$
where $\tilde{\alpha}, \tilde{R}_{m n}, \tilde{\tilde{\xi}}_m$ are $x$ 
independent
($x_m$ being of gradation $0$, $\tilde \alpha, \tilde R_{mn}$
of gradation $1$ and ${\tilde {\tilde \xi}}_m$ of gradation $2$). 
One then has $\delta_\varepsilon x_m =
 \varepsilon^n \tilde{R}_{m n} + \varepsilon_m \tilde{\alpha}, \;\; 
\delta_\varepsilon \tilde{\alpha} = \varepsilon \cdot \tilde{\tilde{\xi}},
\delta_\e \tilde R_{mn} = -(\e_m {\tilde {\tilde \xi}}_n
-\e_n {\tilde {\tilde \xi}}_m) - i \e_{mnpq} \e^p {\tilde {\tilde \xi}}^q$, 
and 
$\delta_\varepsilon \tilde{\tilde{\xi}}_m = \varepsilon_m$. (We do not need
to worry here about the fact that $\delta_\varepsilon$ does not generically 
preserve the reality of $x^m$, $ \tilde{R}_{m n}$ being complex valued, as 
this would just add an extra constraint to the ones we are illustrating here.) 
Let us now try to construct from $X_{+ -}$ a $(+, +)$ multiplet $\Xi_{+ +}
=  \big(\varphi, B,\tA, \ttP,\ttB\big) $.  
This turns out to be extremely constrained. For instance, starting from
an arbitrary function $\ttP \equiv 
\ttP( \tilde{\tilde{\xi}} \cdot \tilde{\tilde{\xi}} )$, one finds that the 
only 
consistent possibility requires
\beqa \label{eqalfa}
&&\tilde{\alpha} = \frac{(\tilde{\tilde{\xi}} \cdot \tilde{\tilde{\xi}})}{2},
 ~~~~~~~~~~~~~~~
\eeqa

\noindent
and reads:
\beqa \label{consteq}
&&\varphi = x \cdot  \tilde{\tilde{\xi} },\;
B_{m n} = constant, \nonumber \\
&&\tilde{A}_m = 
\frac{1}{\sqrt[3]{4}} \; (x_m +  \tilde{\tilde{\xi}}^n  \tilde{R}_{n m} + \alpha 
\tilde{\tilde{\xi}}_m), \nonumber \\
&&\ttB_{m n} = 0, \;
\ttP = \frac{1}{\sqrt[3]{2}} \; \tilde{\tilde{\xi }}\cdot \tilde{\tilde{\xi}}, \nonumber \\
\eeqa

\noindent
 where, furthermore, the 3SUSY transformation of $X_{+ -}$,
 $\delta_\varepsilon X_{+ -}$, induces the 3SUSY transformation of
$\Xi_{+ +}$ but with a specifically rescaled parameter, namely  
$\delta_{ \sqrt[3]{4} \; \varepsilon} \; \Xi_{+ +}$.  
Equations (\ref{eqalfa}, \ref{consteq}) are equally obtained if one starts 
from an arbitrary function $\varphi(x \cdot  \tilde{\tilde{\xi} })$,
[note that (\ref{eqalfa}) is the only relation between $\alpha$
and $\tilde{\tilde{\xi}}$ which is compatible with their transformation
under 3SUSY] . Also similar conclusions
are reached if one starts from $\ttP( \tilde{\tilde{\eta}} \cdot 
\tilde{\tilde{\eta}} )$
or $\varphi(x \cdot  \tilde{\tilde{\eta} })$ where 
$\tilde{\tilde{\eta} } \equiv \tilde{\tilde{\xi}}^n  \tilde{R}_{n m}$, 
or in the case of multi-variable functions.
One even hits impossibilities in more general cases where the components of 
$X_{+ -}$ are assumed to be $x-$dependent. Furthermore, the use of the other
bosonic multiplets or their real-valued combinations (\ref{real}) does not 
improve the 
situation. 

Such strong obstructions are an indication that $x^m$ is not sitting in the 
appropriate multiplets, or equivalently, that a convenient {\sl superspace} 
formulation which weakens as much as possible the constraints has not yet been 
identified.\footnote{It is instructive 
to keep in mind the example of conventional 
supersymmetry. There too, one could have similarly asked whether the 
various scalar, spinor or vector fields should have special space-time 
configurations
in order to belong to the same supermultiplets. For instance, it would 
indeed be so,
in the case of chiral supermultiplets,  
if an inappropriate SUSY transformation of space-time is used. In this context,
the usual superspace formulation can be retrieved from the requirement that
  such potential constraints should be completely relaxed.}
The most natural candidate for such a superspace would make use of the
fermionic 3SUSY (fundamental) multiplets \cite{cubic}. In this case the 
superspace 
would be spanned by $(x^m, \theta_i, \bar{\theta}_i)$, where the $\theta_i$'s 
 ($i=1, 2, 3$)
are $x-$independent anticommuting variables such that
$(\theta_1, \bar{\theta}_2, \theta_3)$ forms a 3SUSY fermionic multiplet which
verifies (\cite{cubic})

\beqa
\label{transfo1}
\delta_\varepsilon \theta_{1  \alpha}& =& \varepsilon^n 
\sigma_{n  \alpha \dot \alpha} \,
\bar{\theta}_2^{\dot \alpha} \nonumber \\
\delta_\varepsilon \bar{\theta}_2^{\dot \beta} & =& \varepsilon^n 
\bar \sigma_{n}^{\dot{\beta} \beta} \, 
\theta_{3  \beta} \\
\delta_\varepsilon \theta_{3 \alpha}& =& 0 \nonumber
\eeqa
 
\noindent
supplemented by the corresponding rules for 
$(\bar{\theta}_1, \theta_2, \bar{\theta}_3)$.
From here the determination of $\delta_\varepsilon x^m$
(which obviously has  to be non-linear) proceeds in a 
well-defined
way, starting from the most general ``cubic superfield" fermionic multiplets
$\psi_i(x^m, \theta_1, \bar{\theta}_1, \theta_2, \bar{\theta}_2, \theta_3, 
\bar{\theta}_3)$ with $i=1, 2, 3$. This lies, however, out of the scope
of the present paper,  and will be treated elsewhere.

\section{Summary and outlook}
\renewcommand{\theequation}{4.\arabic{equation}}   
\setcounter{equation}{0}
In this paper, we have continued the study of cubic supersymmetry
initiated in \cite{cubic}. Here, we focused  on 
the bosonic multiplets leaving aside the fermionic ones.
We considered the most general 3SUSY invariant
Lagrangian which is quadratic in the fields, exhibited  explicitly 
its diagonal form and argued for a possible solution for the unboundedness 
from below of the energy density encountered in \cite{cubic}.
Furthermore, we studied interaction terms involving {\it only}
the bosonic multiplets and  proved that 3SUSY  forbids such terms altogether. 
Such an obstruction strengthens the interpretation of the boson multiplets
in terms of abelian gauge fields for which ``renormalizable"
self-interactions are absent and where the gauge is fixed {\sl \`a la }
Feynman. We also looked in some detail at the residual gauge symmetry
and the possibility of identifying the physical degrees of freedom in this
context. A related question emerged as to whether the 3SUSY algebra would
imply rather specific space-time configurations for the partner fields
(in contrast with the case of conventional supersymmetry). An unambiguous 
answer
to this question requires the identification of the proper 3SUSY transformation
of space-time, for which we only sketched a superspace approach in this paper.
The more general question regarding the possibility of an interacting
theory would still have to be further investigated. Among the possible
directions one could consider coupling the bosonic multiplets to fermionic 
ones, [albeit highly non conventional kinetic terms for the latter 
\cite{cubic}],
or more general boson multiplets bi-linear in fermionic fields which are
all charged under 3SUSY. One can also consider extended 3SUSY algebras
with $N$ copies of the $Q$ generators offering the possibility that the
associated automorphism group would induce non-abelian structures.
Such a possibility would involve non-abelian self-interacting $p$-forms which
of course require a careful investigation given 
the strong constraints when $p \ge 2$ (see {\it e.g}  \cite{hk}).     

Other types of extensions of the Poincar\'e algebra, namely parasupersymmetric
extension  have been considered in \cite{psusy}.
A natural question one might address is the relation between these
two extensions. Since parasupersymmetry admits interaction terms,
this  relation (if it  exists)  could give some indication
on the interaction possibilities for 3SUSY. 

Finally, a perhaps more promising approach would make use of the fact that
3SUSY has a natural extension in an arbitrary number of space-time dimensions
\cite{mrt}, and that the $p-$forms of the bosonic 3SUSY multiplets couple 
naturally to extended objects of dimension $(p-1)$ ($(p-1)-$branes). 
The proper 
transformations of these extended objects are then determined by the 3SUSY
transformation of space-time and one can seek for a 3SUSY invariant theory
for interacting $p$-branes.

\bigskip
{\bf Ackowledgement}: We would like to acknowledge J. Lukierski 
for useful discussions and
remarks.

\end{document}